\documentclass[aps,prx,twocolumn,longbibliography]{revtex4-1}

\usepackage{amsmath}
\usepackage{amssymb}
\usepackage{graphicx}
\usepackage{bm,hyperref}
\usepackage{braket}
\usepackage{xcolor}
\usepackage{array}
\usepackage{bbold}
\newcolumntype{C}[1]{>{\centering\arraybackslash}p{#1}}

\renewcommand{\Re}{\textrm{Re}}
\renewcommand{\Im}{\textrm{Im}}
\newcommand{\tr}[1]{\text{tr}\left[{#1}\right]}

\def\vec{\mathbf}
\def\ovarrow{\overrightarrow}

\begin{document}
\title{In-plane orbital magnetization as a probe for symmetry breaking in strained twisted bilayer graphene}

\date{\today}
\author{Ohad Antebi}
\affiliation{Department of Condensed Matter Physics, Weizmann Institute of Science, Rehovot 76100, Israel}
\author{Ady Stern}
\affiliation{Department of Condensed Matter Physics, Weizmann Institute of Science, Rehovot 76100, Israel}
\author{Erez Berg}
\affiliation{Department of Condensed Matter Physics, Weizmann Institute of Science, Rehovot 76100, Israel}
\begin{abstract}

Three symmetries prevent a twisted bilayer of graphene from developing an in-plane spontaneous magnetization in the absence of a magnetic field - time reversal symmetry,  $C_2$ symmetry to $\pi$ rotation and $C_3$ symmetry to $2\pi/3$ rotation. In contrast, there are experimental and theoretical indications that, at certain electron densities, time reversal and $C_2$ symmetries are broken spontaneously, while the $C_3$ symmetry is often broken due to strain. We show that in-plane orbital magnetization is a very sensitive probe to the simultaneous breaking of these three symmetries, exhibiting surprisingly large values (of the order of one Bohr magneton per moir\'{e} unit cell) for valley polarized states at rather small values of heterostrain. We attribute these large values to the large magnitude of the characteristic magnetization of individual Bloch states, which we find to reflect the fast Dirac  velocity of single layer graphene, rather than the slow velocity of the twisted bilayer. These large values are forced to mutually cancel in valley-symmetric states, but the cancellation does not occur in valley-polarized states. Our analysis is carried out both for non-interacting electrons and within a simplified Hartree-Fock framework. 

\end{abstract}
\maketitle

\section{Introduction}

Stacking two graphene layers with a relative twist near the ``magic angle''\cite{Bistritzer2011} has produced a flurry of unexpected physical phenomena\cite{Cao2018,Cao2018a,Yankowitz2019,Arora2020,Choi2019,Wong2020,Saito,Stepanov2020,Uri2020,Xie2019,Rozen2021,Sharpe2019,Park2021,Jiang2019,Kerelsky2019,Lu2019,Zondiner2020,Serlin2020,Cao2021,Tschirhart2021,Saito2021,Rozen2021}, including a variety of symmetry broken phases\cite{Jiang2019,Kerelsky2019,Lu2019,Zondiner2020,Serlin2020,Cao2021,Tschirhart2021,Saito2021,Rozen2021}, such as spin and valley polarized states\cite{Sharpe2019,Zondiner2020,Park2021}.
An important question is how are these patterns of broken symmetry manifested in measurable physical quantities. 
At first glance, it seems that although an in-plane magnetic field is a good mean to probe spin polarization, it is an unlikely candidate for probing \emph{orbital} response in twisted bilayer graphene (TBG). This is so because the physical separation between the two layers is of atomic scale, and at attainable magnetic fields, the flux through the cross sectional area of a unit cell is a small fraction of a flux quantum. 

In this work, we find that, in fact, the single-electron Bloch states of TBG generically possess a large in-plane orbital magnetization, of the order of several Bohr magnetons. We trace this observation to the nature of these Bloch states, which consist of superpositions of components localized at the two layers. These components move at the Dirac speed of a single graphene layer, but their directions of motion are nearly opposite, resulting in the slow group velocity of the narrow TBG bands. This state of affairs is reminiscent of the familiar ``zitterbewegung'' phenomenon of relativistic massive Dirac electrons \cite{Breit1928}, which jitter rapidly close to the speed of light even when their average velocity is non-relativistic. In TBG, the rapid inter-layer relative motion produces a large in-plane orbital magnetization. 

Despite the large magnetic moment of the individual Bloch states, the net in-plane magnetization vanishes if the system possesses either time reversal or $C_3$ rotational symmetry. However, both of these symmetries are frequently broken in TBG, either externally or spontaneously. The $C_3$ symmetry can be broken when the system is subjected to strain. Time reversal symmetry can be broken spontaneously due to electron-electron interactions. Both forms of symmetry breaking were observed in experiments \cite{Choi2019,Kerelsky2019,Xie2019,Kazmierczak,Jiang2019,Sharpe2019,Serlin2020,Tschirhart2021}. Here, we show that in the presence of inter-layer strain as small as $\sim 10^{-4}$, a spontaneously valley polarized state can have an in-plane magnetic moment as large as a few $\mu_B$ per moir\'e unit cell. The in-plane magnetization can be extracted from measurements of the electronic compressibility in an in-plane magnetic field, by using thermodynamic relations \cite{Zondiner2020}. 

Our analysis consists of several parts. In the first part, we compute the in-plane orbital magnetization of individual Bloch states within the Bistritzer-MacDonald (BM) model. We then show that in the presence of an inter-layer strain, the total in-plane moment of each valley is non-zero, and calculate the moment as a function of the magnitude and direction of the strain. Finally, we consider the effect of electron-electron interactions within a simplified Hartree-Fock analysis, and show that if the interactions lead to the breaking of the valley symmetry, the resulting many-body state has a large in-plane magnetic moment.  

This paper is organized as follows: Section \ref{sec:setup} describes the non-interacting Hamiltonian in the presence of strain and an in-plane magnetic field. Section \ref{sec:general_considerations} presents analytical results derived from symmetry considerations and illustrates them in a toy model. Section \ref{sec:numerical} presents numerical results for the non-interacting model, and presents a simplified interacting model and corresponding results. Section \ref{sec:discussion} concludes the paper.

\section{Setup}\label{sec:setup}

In this section we present a non-interacting model for strained twisted bilayer graphene with in-plane magnetic fields, which is a generalization of the BM Hamiltonian. For a pedagogical construction of the effects of strain on the BM Hamiltonian we refer readers to Refs.~[\onlinecite{Bi2019,balents2019general}], and give the result here for completeness and in order to set the notation.

We consider the following continuum Hamiltonian for valley $K$ in the layer basis:
\begin{equation}
    H_{\text{BM},K} = \begin{pmatrix}
    h^{(1)}(\vec{k}) & T(\vec{r}) \\
    T^{\dagger}(\vec{r}) & h^{(2)}(\vec{k})
    \end{pmatrix},
\end{equation}
where $h^{(1)},h^{(2)}$ are the single layer Hamiltonians of the deformed graphene sheets near valley $K$, and $T(\mathbf{r})$ is the interlayer coupling term (see below). The opposite valley's Hamiltonian $H_{\text{BM},K'}$ is obtained by time-reversal, and is assumed to be decoupled from $H_{\text{BM},K}$. The momentum space representation of the continuum intralayer Hamiltonian in layer $l$ is given to linear order in the (unsymmetrized) strain tensor as follows:
\begin{equation}
    h^{(l)}(\vec{k}) = \hbar v_{F} \left[(\mathbb{1}+\mathcal{U}^{(l)})^{T}(\vec{k}-\delta\vec{K}^{(l)})\right]
    \cdot\begin{pmatrix}\sigma_{x},\sigma_{y}\end{pmatrix}.
\end{equation}
Here $v_{F}=0.9\times 10^{6}\frac{m}{s}$ is the bare Dirac velocity, $\vec{k}$ is measured with respect to the $K$ point of the undeformed graphene layer. $\delta\vec{K}^{(l)}$ is the shift in position of the Dirac node of layer $l$ from its undeformed position and will be elaborated upon shortly. The matrix $\mathcal{U}^{(l)}$ is the unsymmetrized strain tensor whose components are $u_{ij}^{(l)}=\partial_{i}u_{j}^{(l)}$, with $u_{j}^{(l)}(\mathbf{r})$ being the displacement field along the $j$ axis. The antisymmetric and symmetric components of this strain tensor will be hereby denoted as:
\begin{equation}\label{eq:strain_tensor_sym_asym}
    \mathcal{U}^{(l)}
    =
    \begin{pmatrix}
    s_{xx}^{(l)} & s_{xy}^{(l)}-\theta^{(l)} \\
    s_{xy}^{(l)}+\theta^{(l)} & s_{yy}^{(l)}
    \end{pmatrix}
    =
    \mathcal{R}(\theta^{(l)})+\mathcal{S}^{(l)},
\end{equation}
where the antisymmetric part $\mathcal{R}(\theta^{(l)})$ describes a rotation of the layer, and the symmetric part $\mathcal{S}^{(l)}$ includes the traceless strain that breaks $C_3$ symmetry. We twist the layer symmetrically by choosing $\theta^{(1)}=-\theta^{(2)}=\theta/2$. The discussion of the  strain tensor is restricted to systems with $\mathcal{S}^{(1)}=\chi\mathcal{S}^{(2)}$ where $\chi=-1$ ($\chi=+1$) describes heterostrain (homostrain). It is assumed throughout that the strain is homogeneous and weak compared to the twist angle, i.e. $s_{ij}\ll\theta \ll 1$.

The position of the Dirac node in a deformed graphene monolayer is determined by a combination of geometric and intra-layer energetics considerations. Geometric considerations cause the original Dirac node to be rotated by the twist and then shifted by the strain, summarized as $\vec{K}\rightarrow(1-\mathcal{U}^{(l)})^{T}\vec{K}$. Here $\vec{K}=\left(\frac{4\pi}{3\sqrt{3}a},0\right)$ is the momentum of the original Dirac node, and $a=0.142$nm is the distance between carbon atoms. By accounting for the effect of strain on the intralayer hopping, the Dirac node is shifted further, an effect that can be described using an artificial gauge field \cite{Suzuura2002,CastroNeto2009}:
\begin{equation}\label{eq:artificial_gauge_field}
    \vec{A}_{s}^{(l)} = \frac{\beta}{2a}\begin{pmatrix}
    s_{xx}^{(l)}-s_{yy}^{(l)},-2s_{xy}^{(l)}
    \end{pmatrix},
\end{equation}
with $\beta\approx3.14$ being a dimensionless parameter \cite{Nam2017}. Note that the gauge field couples oppositely to the two valleys, and hence does not break time-reversal symmetry.
The combined shift in position of the Dirac node due to twist and strain, $\delta\vec{K}^{(l)}$, is therefore given by:
\begin{align}\label{eq:dirac_node_position_strain}
\begin{split}
    \delta\vec{K}^{(l)}&=-(\mathcal{U}^{(l)})^{T}\vec{K}-\vec{A}_{s}^{(l)}\\
    &=(\mathcal{R}(\theta^{(l)})-\mathcal{S}^{(l)})\vec{K}-\vec{A}_{s}^{(l)}.
\end{split}
\end{align}
Note that, in the absence of strain, this shift is simply $\delta\vec{K}^{(l)}=-\theta^{(l)}\hat{z}\times\vec{K}$.

The tunneling amplitude between the layers will have the periodicity of the strained moir\'{e} pattern. We allow for different tunneling amplitudes between AA and AB sites, denoted by $t_{AA}$ and $t_{AB}$ respectively. Accounting for lattice relaxation \cite{Koshino2018,Carr2019}, we set $t_{AA}=0.7t_{AB}$ with $t_{AB}=0.11\,\rm{eV}$. 
The tunneling in the sublattice basis can be written as:
\begin{equation}
    T(\vec{r}) = 
    \sum_{j=0,1,2} T_{j}
    e^{-i\vec{g}_{j}\cdot\vec{r}},
\end{equation}
where $T_{j} = t_{AA}+t_{AB}\left[\Re{(\zeta^{j})}\sigma_{x}-\Im{(\zeta^{j})}\sigma_{y}\right]$, and $\zeta=e^{i2\pi/3}$.
The moir\'{e} reciprocal lattice vectors are given by $\vec{g}_{j}=\mathcal{U}^{T}\vec{G}_{j}$ where $\mathcal{U}=\mathcal{U}^{(1)}-\mathcal{U}^{(2)}$ is the relative twist and strain, and $\vec{G}_{0}=0,\vec{G}_{1}=\sqrt{3}K\left(\frac{\sqrt{3}}{2},\frac{1}{2}\right),\vec{G}_{2}=\sqrt{3}K\left(\frac{\sqrt{3}}{2},-\frac{1}{2}\right)$ are the graphene reciprocal lattice vectors. While the Dirac node moves for either homostrain or heterostrain, to leading order in the strain and twist angle only heterostrain modifies the moir\'{e} pattern and hence the tunneling between the layers.

Next, we consider the application of an in-plane magnetic field. A uniform in-plane field $\vec{B}_{\parallel}=|B|(\cos{\phi_{B}},\sin{\phi_{B}})$ can be described by a $z$-dependent, in-plane vector potential $\vec{A}(z) = z|B|(\sin{\phi_{B}},-\cos{\phi_{B}})$. We choose $z=0$ to be halfway between the layers, which are separated by distance $d=0.335\,\rm{nm}$, leading to constant but opposite gauge fields for different layers. Defining  
\begin{equation}
    \vec{A}_{\text{em}} \equiv \vec{A}\left(z=\frac{d}{2}\right) = \frac{1}{2}|B|d(\sin{\phi_{B}},-\cos{\phi_{B}}),
\end{equation}
we can minimally couple the in-plane magnetic field to the TBG as follows:
\begin{equation}\label{eq:h_bm_w_mag}
    H_{\text{BM},K} = \begin{pmatrix}
    h^{(1)}\left(\vec{k}-\frac{e}{\hbar}\vec{A}_{\text{em}}\right) & T(\vec{r}) \\
    T^{\dagger}(\vec{r}) & h^{(2)}\left(\vec{k}+\frac{e}{\hbar}\vec{A}_{\text{em}}\right)
    \end{pmatrix}
    -\bm{\mu}_{s}\cdot\vec{B}_{\parallel},
\end{equation}
where $\bm{\mu}_{s}$ is the electron's spin magnetic moment.
Note that the local tunneling amplitude is unaffected by the in-plane field. This completes our non-interacting description of the strained TBG with in-plane magnetic field.

\section{General considerations}\label{sec:general_considerations}
\subsection{Symmetry analysis}
The zero-temperature, zero-field in-plane magnetization per moir\'{e} unit cell is defined as:
\begin{equation}\label{eq:inplane_magnetization_def}
    \vec{M}_{\parallel} = -\left.\frac{1}{N_{u.c.}}\frac{\partial E_{G.S.}}{\partial \vec{B}_{\parallel}}\right|_{\vec{B}_{\parallel}=0}
\end{equation}
where $E_{G.S.}=\bra{\Psi_{G.S.}} H \ket{\Psi_{G.S.}}$ is the energy of the ground state $\ket{\Psi_{G.S.}}$, and $N_{u.c.}$ is the number of moir\'{e} unit cells.
We are interested in the restrictions symmetry imposes on this magnetization. 
We will show that both valley polarization and traceless heterostrain are necessary for a non-vanishing orbital magnetization. Furthermore, to linear order in strain, the magnitude of the magnetization is independent of the strain direction.

We start by listing the symmetries of the unstrained TBG, in the absence of any magnetic field. The symmetries of a single unstrained graphene layer are given by two and three-fold rotations about the $z$ axis, $C_2$, $C_3$, two mirror symmetries $M_{y,z}$ about the $x-z$ and $x-y$ planes, respectively, time-reversal $\mathcal{T}$, and $SU(2)$ spin rotation symmetry. The twisted bilayer as a whole breaks $M_y$ and $M_z$, but has two additional symmetries given by $R_{\pi}^{x}$ and $R_{\pi}^{y}$, which are $\pi$-rotations about the $x$ and $y$ axes, respectively, in the plane halfway between the layers (note that these symmetries interchange the two layers).

In order to avoid an overly complicated description, we restrict our attention to a subset of strains and flavor symmetry breaking configurations of interest. We restrict the discussion of strain to that of either heterostrain or homostrain, i.e. $\mathcal{S}^{(1)}=\chi\mathcal{S}^{(2)}\equiv \mathcal{S}$ with $\chi=-1$ or $\chi=+1$, respectively. For the flavor symmetry, we anticipate any $C_2$ or $\mathcal{T}$ symmetric state to yield $\vec{M}_{\parallel}=0$. To highlight orbital magnetization we consider variable fillings of the different valleys, $\lbrace \nu_{K},\nu_{K'}\rbrace$, with $\nu_{K(K')}=\nu_{K(K')\uparrow}+\nu_{K(K')\downarrow}$ (where $\uparrow$, $\downarrow$ indicate the two spin directions). Equipped with these parameters, we write the ground state energy as $E\left(\mathcal{S},\vec{B}_{\parallel},\nu_{K},\nu_{K'}\right)$, and explore how the different symmetries constrain this energy.

The $C_2$ symmetry constrains the energy to the form:
\begin{equation}\label{eq:energy_symmetry_c2}
    E\left(\mathcal{S},\vec{B}_{\parallel},\nu_{K},\nu_{K'}\right)
    =E\left(\mathcal{S},-\vec{B}_{\parallel},\nu_{K'},\nu_{K}\right).
\end{equation}
We note that this symmetry flips the valleys and the magnetic field, but does not affect the strain tensor. Time reversal symmetry imposes the same constraint. By examining \eqref{eq:inplane_magnetization_def} and \eqref{eq:energy_symmetry_c2} it is clear that finite in-plane magnetization requires flavor symmetry breaking in the form of valley polarization, i.e. $\nu_{K}\neq \nu_{K'}$. 

Next, we consider the effect of $C_3$ rotation. This symmetry preserves the valley fillings, but rotates both magnetic field and strain:
\begin{equation}\label{eq:energy_symmetry_c3}
    E\left(\mathcal{S},\vec{B}_{\parallel},\nu_{K},\nu_{K'}\right)
    =E\left(C_3\mathcal{S}C^T_3,C_3\vec{B}_{\parallel},\nu_{K},\nu_{K'}\right).
\end{equation}
with $C_3=\exp{\left(i2\pi\eta_{y}/3\right)}$ and $\eta_{y}$ is the $y$ Pauli matrix. Expanding \eqref{eq:energy_symmetry_c3} in components of $\mathcal{S}$ and $\vec{B}_{\parallel}$ introduces different constraints on different powers of strain and magnetic field. The zero-field magnetization is given by the coefficient of the first order in magnetic field. By considering the $\mathcal{S}=0$ case, we immediately see that zero field in-plane magnetization is prohibited for an unstrained TBG, regardless of flavor symmetry breaking, due to the $C_3$ symmetry. 

We wish to construct a magnetization vector that is linear in strain elements and rotates under $C_3$ as dictated by Eq.~\eqref{eq:energy_symmetry_c3}.
When considering rotations of a strain tensor it is most natural to think of rotating its principal axis.
We therefore continue with the diagonalized form of the strain tensor: 
\begin{equation}\label{eq:strain_tensor_prinicpal_axis}
    \mathcal{S}=
    \begin{pmatrix}
    \cos{\phi_{S}} & -\sin{\phi_{S}} \\
    \sin{\phi_{S}} & \cos{\phi_{S}}
    \end{pmatrix}
    \begin{pmatrix}
    s' & 0 \\
    0 & s''
    \end{pmatrix}
    \begin{pmatrix}
    \cos{\phi_{S}} & \sin{\phi_{S}} \\
    -\sin{\phi_{S}} & \cos{\phi_{S}}
    \end{pmatrix}.
\end{equation}
where $\phi_{S}$ is the angle between the principal axis and the $x$ axis, and $s',s''$ denote the principal strains. The $C_3$ rotation of the strain tensor only transforms $\phi_{S}\rightarrow\phi_{S}+2\pi/3$, leaving the principal strains invariant.

Any rotationally invariant component of the strain tensor would not contribute to the magnetization vector at linear order. 
The strain tensor in \eqref{eq:strain_tensor_prinicpal_axis} can be split into the dilatation and the traceless part as
\begin{equation}\label{eq:dilatation_and_traceless}
        \mathcal{S}=\frac{s'+s''}{2}\mathbb{1}+\frac{s'-s''}{2}(\eta_x\sin2\phi_S  + \eta_z\cos2\phi_S ),
\end{equation}
where we denote the identity matrix by $\mathbb{1}$, and $\eta_{x,z}$ are Pauli matrices. 
Clearly, the dilatation is rotationally invariant, and therefore does not contribute to the magnetization at linear order, leaving only the traceless part at this order. We conclude that the magnetization is proportional to $s'-s''$, the difference between principal strains. 
Since the angle $2\phi_{S}$ transforms under $C_3$ rotation as $2\phi_{S}\rightarrow2\phi_{S}+4\pi/3$, while the magnetization rotates by $2\pi/3$, we find that the following form of the energy satisfies Eq.~\eqref{eq:energy_symmetry_c3}: 
\begin{align}\label{eq:magnetization_from_c3}
\begin{split}
    &E\left(\mathcal{S},\vec{B},\nu_{K},\nu_{K'}\right) \cong
    -\vec{M}_{\parallel}\cdot\vec{B}_{\parallel}
    \\
    &\cong \lambda(\nu_{K},\nu_{K'}) |B|(s'-s'')\cos{(2\phi_{S}+\phi_{B}+\phi_{0})},
\end{split}
\end{align}
where $\lambda(\nu_{K},\nu_{K'})=-\lambda(\nu_{K'},\nu_{K})$ is an odd function under exchange of its arguments, $\phi_{0}\in[0,\pi)$ is a constant angle soon to be determined, and $\vec{B}_{\parallel}=|B|\left(\cos{\phi_{B}}, \sin{\phi_{B}}\right)$. Note that the magnitude of the magnetization is independent of the strain's principal axis at linear order. This observation does not hold to higher orders in strain, as shown in detail in Appendix~\ref{sec:app_symmetry}. 

Eq.~\eqref{eq:magnetization_from_c3} clearly shows how the magnetization respects the $C_3$ symmetry. We still, however, have a degree of freedom in choosing $\phi_{0}$. This phase will turn out to depend on whether the system is homostrained or heterostrained, and it is determined by symmetries that relate the two graphene layers.

Using the $R_{\pi}^{x}$ symmetry which preserves the valley fillings, we find the energy must be symmetric under the transformation $\mathcal{S}\rightarrow\chi\mathcal{S}$ (due to layer flip), $\phi_{B}\rightarrow-\phi_{B}$ and $\phi_{S}\rightarrow-\phi_{S}$. For heterostrain ($\chi=-1$), this constraint enforces $\phi_{0}=\pi/2$, yielding the following form for the magnetization energy:
\begin{equation}
    \vec{M}_{\parallel}\cdot\vec{B}_{\parallel}\cong \lambda(\nu_{K},\nu_{K'}) |B|(s'-s'')\sin{(2\phi_{S}+\phi_{B})}.
\end{equation}
For the homostrained case we obtain $\phi_0=0$ and one can replace the sine above by a cosine. It can be verified that the symmetry $R_{\pi}^{y}$ does not add any other constraint (see Appendix~\ref{sec:app_symmetry}). The magnetization immediately follows, and is given here for completeness:
\begin{equation}
    \vec{M}_{\parallel,x} + i\vec{M}_{\parallel,y}\cong \lambda(\nu_{K},\nu_{K'})(s'-s'')
    e^{-i\left(2\phi_{S}+\phi_{0}\right)},
    \label{eq:M}
\end{equation}
with $\phi_0 = \pi/2$ ($\phi_0 = 0$) for heterostrain (homostrain), respectively. 
Motivated by the result above, we will proceed by considering only traceless strain tensors $\mathcal{S}$ with $s'=-s''$ such that the leading order contribution does not vanish. 

\subsection{Qualitative discussion of orbital magnetization}\label{sec:qualitative}

The in-plane orbital magnetic moment of a Bloch state can be pictorially described as resulting from a current loop consisting of components in opposite layers, moving in opposite directions close to the single-layer Dirac velocity $v_F$. 
Close to the magic angle, the net group velocity of the Bloch state is much smaller than $v_F$. 
We will show, however, that the flatness of the resulting bands is still important for the total magnetization in the presence of strain. As shown in Refs.~\onlinecite{Huder2018,Qiao2018,Bi2019,Liu2021nematic,Parker2021}, the flat bands are extremely susceptible to strain, thus allowing for a significant rotation symmetry breaking near the magic angle.

With this picture in mind, we estimate the magnetic moment of a Bloch state by constructing a current loop from the physical parameters of the bilayer. 
The current in the loop is estimated to be $I\sim ev_{F}/L$, where $L$ is a length scale that will soon cancel out. The loop area is given by the same length scale $L$ times the distance between the layers $d$. We therefore find that orbital magnetic moment associated with such a loop is given by $|M_{\parallel}|\sim ev_{F}d\sim 5\mu_{B}$, where $\mu_{B}$ is the Bohr magneton. We show below that this overestimates the magnetization by a factor of $2-3$.

The simplest model that allows for a quantitative analytical calculation of the magnetization is the ``Tripod model"\cite{Bistritzer2011,Bernevig2021}. Within this model, the effective Hamiltonian with applied heterostrain and in-plane magnetic field is given by:
\begin{align}
\begin{split}
    H(\vec{k}) &= h^{(1)}\left(\vec{k}-\frac{e}{\hbar}\vec{A}_{\text{em}}\right)\\
    &-\sum_{j=0,1,2}T_{j}
    \left[
    h^{(2)}\left(\vec{k}+\frac{e}{\hbar}\vec{A}_{\text{em}}+\vec{g}_{j}\right)-E(\vec{k})
    \right]^{-1}T_{j}^{\dagger},
\end{split}
\label{eq:tripod}
\end{align}
where the energy is to be solved self-consistently using $H(\vec{k})\ket{\Psi(\vec{k})}=E(\vec{k})\ket{\Psi(\vec{k})}$. This model is valid in the vicinity of layer $1$'s Dirac point, i.e. where $\vec{p}=\vec{k}-\delta\vec{K}^{(1)}$ is small compared to the moir\'{e} BZ. We expand this model to first order in $p/(K\theta)$ and in strain in Appendix~\ref{sec:app_tripod}. Omitting a $\vec{p}$-independent constant, the result for valley $\tau=K$ is given by:
\begin{equation}\label{eq:tripod_to_first_order}
    H(\delta\vec{K}^{(1)}+\vec{p}) \cong \hbar v_{F}\left(\frac{v_{F}^{*}}{v_{F}}(\vec{p}-\vec{p}_{0})-2\frac{v_{F,A}}{v_{F}}\frac{e}{\hbar}\vec{A}_{\text{em}}\right)\cdot\vec{\sigma},
\end{equation}
with the renormalized velocities given by:
\begin{align}\label{eq:renormalized_velocities}
\begin{split}
    \frac{v_{F}^{*}}{v_{F}} &= \frac{1-3w^2}{1+3(u^2+w^2)},\\
    \frac{v_{F,A}}{v_{F}} &= \frac{1}{1+3(u^2+w^2)},
\end{split}
\end{align}
where $u=t_{AA}/(\hbar v_{F}K\theta)$ and $w=t_{AB}/(\hbar v_{F}K\theta)$ are the dimensionless tunneling parameters.
 
The tripod model predicts a shift of the Dirac points in the presence of strain due to the breaking of $C_3$ symmetry, given by:
\begin{equation}\label{eq:tripod_shift_by_strain}
    \vec{p}_{0} = \frac{3w^2}{1-3w^2}\left(1+\frac{u^2}{w^2}\frac{Ka}{\beta}\right)\vec{A}_{s},
\end{equation}
where $\vec{A}_{s}=\vec{A}_{s}^{(1)}-\vec{A}_{s}^{(2)}$, as given by \eqref{eq:artificial_gauge_field}, and is proportional to the strain. Note that for homostrain $\vec{A}_{s}$ vanishes, and for heterostrain it does not. The reason is that, to first order in strain, only the strain difference between layers modifies the moir\'{e} pattern.

We define the magic angle condition to be $v_{F}^{*}=0$, or equivalently $w^2=1/3$. By examining Eq.~\eqref{eq:tripod_shift_by_strain} we find, for a heterostrained TBG, an acute sensitivity of the Dirac point position to the twist angle in the vicinity of the magic angle condition. A similar strong sensitivity of the Dirac points to twist angle in the presence of an in-plane magnetic field was discussed in Ref. \cite{Kwan2020}. The acute sensitivity is manifest also in the full numerical calculation which extends beyond the tripod model. The large shift of the Dirac point indicates a strong breaking of $C_{3}$ symmetry, and therefore allows for a substantial net in-plane orbital magnetization per moir\'{e} unit cell, as section \ref{sec:numerical} will show. 

Consider the magnetization of a particular momentum state:
\begin{equation}
    \vec{M}_{\parallel}(\vec{p}) = -\left.\frac{\partial E_{-}(\vec{p})}{\partial \vec{B}_{\parallel}}\right|_{\vec{B}_{\parallel}=0}
\end{equation}
where $E_{-}(\vec{p})$ is the lower energy band. 
The in-plane magnetization is readily calculated from Eq.~\eqref{eq:tripod_to_first_order} to be
\begin{equation}
    \vec{M}_{\parallel}(\vec{p}) = e v_{F,A} d \frac{\hat{z}\times (\vec{p}-\vec{p}_{0})}{\left|\vec{p}-\vec{p}_{0}\right|}.
\end{equation}
This result is in agreement with our initial discussion of the expected magnetization magnitude, up to a renormalization of the velocity due to tunneling. The beauty of this result is that, indeed, the renormalized velocity $v_{F,A}$ that enters the magnetization does not vanish for any twist angle.
At the magic angle, our new estimate for the magnetization magnitude is given by $|M_{\parallel}|=e v_{F,A} d\approx 2\mu_{B}$. We will see that this result is a fair estimate for the total magnetization per moir\'{e} unit cell when the $C_{3}$ rotation and time-reversal $\mathcal{T}$ symmetries are strongly broken.

\subsection{Effects of interactions}

The vanishing Dirac velocity at the magic angle is a property of the non-interacting band structure. By including electron-electron interactions, one expects the Dirac velocity and the bandwidth to modified. Therefore, a natural question to ask is - How do interactions affect the strain sensitivity of the $C_{3}$ symmetry breaking near the magic angle? We will demonstrate in section~\ref{sec:numerical} that, within our simplified Hartree-Fock treatment, the necessary strain for a significant total magnetization remains within reasonable experimental values. We will also show that the magnetization magnitudes of individual Bloch states are practically unchanged by the interactions, due to the insensitivity to interactions of the parameters that determine it, $v_{F,A}$ and $d$.

\section{Numerical analysis}\label{sec:numerical}

\subsection{Valley resolved magnetization in a non-interacting model}
The non-interacting ground state, even in the presence of strain, has a zero overall orbital magnetization due to time-reversal symmetry. However, it is still instructive to calculate the contribution of a single valley to the net orbital magnetization. This calculation will become useful when we discuss states which are valley polarized due to interactions.

In the non-interacting model, the ground state energy is the sum of energies of single particle Bloch states, $E_{G.S.}=\sum_{\alpha,\vec{k}}^{\prime}E_{\alpha}(\vec{k})$, where $\alpha=(\tau,s,m)$ is a combined label for valley $\tau$, spin $s$ and band index $m$, and the prime restricts the summation to occupied states. Similarly, the total magnetization in Eq.~\eqref{eq:inplane_magnetization_def} is given by an average over the magnetizations of individual Bloch states: $\vec{M}_{\parallel}=\sum_{\alpha,\vec{k}}^{\prime}\vec{m}_{\alpha}(\vec{k})/N_{u.c.}$. 

\begin{figure*}
    \centering
    \includegraphics[width=7.0in]{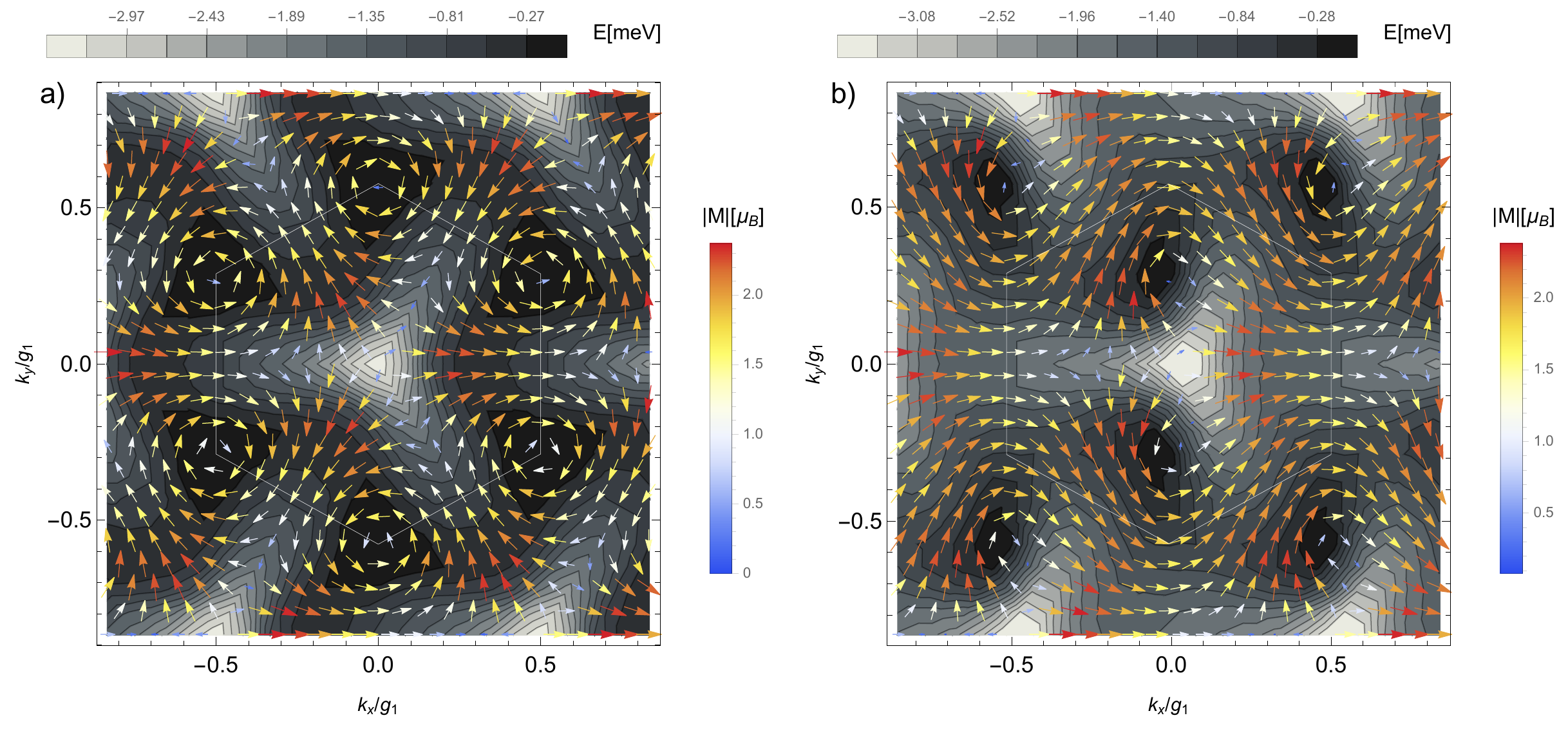}
    \caption{Orbital in-plane magnetization vectors of individual Bloch states in the valence band of valley $\tau=K$, plotted across the moir\'{e} BZ for the noninteracting TBG with twist angle of $\theta=1.1^{\circ}$. Background contours represent the energy band. a) No strain, b) Heterostrain with principal strains $s'=-s''=2\times10^{-4}$ and angle $\phi_{S}=\pi/4$. Notice how the magnetization of each state is $\sim 1\mu_{B}$ for both cases, but in the unstrained case  the total magnetization vanishes due to $C_{3}$ symmetry. The hexagon outlines the moir\'{e} BZ, which is effectively unchanged by the strain for $s'/\theta \sim 10^{-2}$.}
    \label{fig:mag_vs_bz_nonint}
\end{figure*}

The contribution of a single band and flavor combination $\alpha$ to the magnetization can be visualized by plotting the orbital component (ignoring the trivial contribution of spin moments) of $\vec{m}_{\alpha}(\vec{k})$ in the entire moir\'{e} Brillouin zone. 
Focusing on the valence band (the lower flat band) of valley $K$, we plot in Fig.~\ref{fig:mag_vs_bz_nonint} the magnetization for strain-free and heterostrained MATBG. We find that the orbital magnetization of each Bloch state is of order $\sim1\mu_{B}$ in both cases, and the contribution of the entire band to the magnetization is determined by the strength of the $C_3$ symmetry breaking due to the strain. 

\begin{figure}[b]
    \centering
    \includegraphics[width=0.5\textwidth]{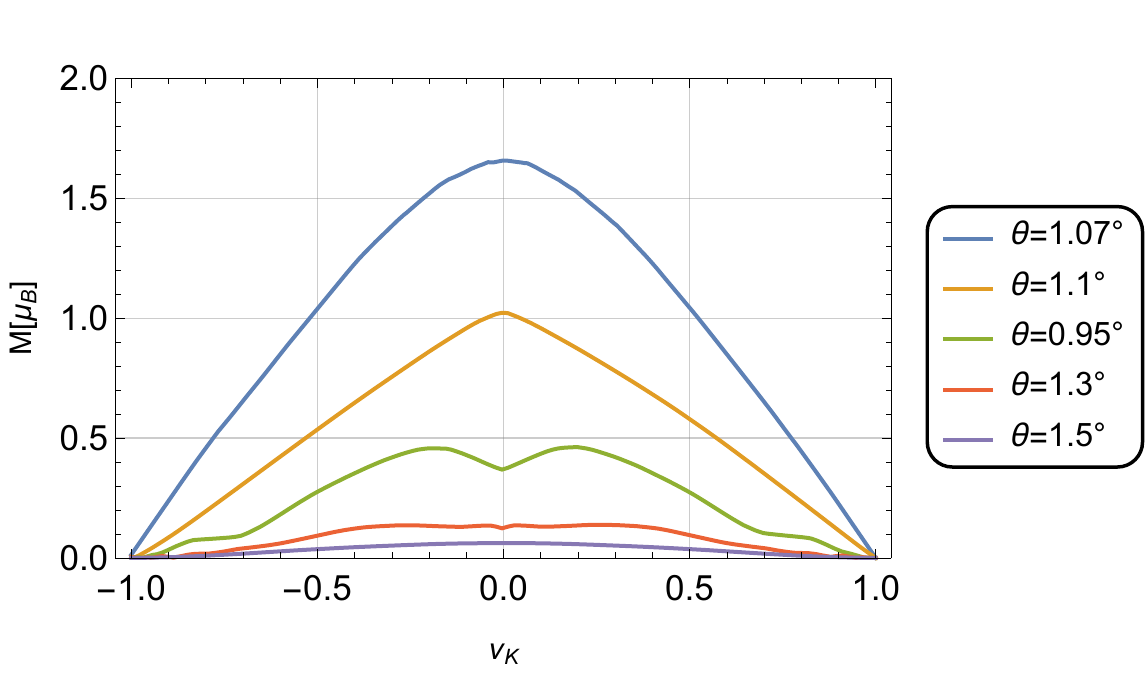}
    \caption{The valley-$K$ resolved magnitude of the magnetization \textit{vs} filling for the noninteracting TBG with various twist angles. The filling $\nu_{K}$ denotes the number of electrons per moir\'{e} unit cell with respect to charge neutrality. 
    The strain parameters are specified in Fig.~\ref{fig:mag_vs_bz_nonint}b.
    Note the strong magnetization near the magic angle $\theta_{MA}\approx 1.07^{\circ}$, and that the magnetization of the valence band is exactly cancelled by that of the conduction band. }
    \label{fig:mag_vs_filling_nonint}
\end{figure}

Due to the qualitative considerations of Sec.~\ref{sec:qualitative}, we expect the strength of the symmetry breaking to be enhanced in the vicinity of a magic angle. We confirm this enhancement by computing the valley resolved magnetization for different twist angles as a function of filling.
We will restrict our sum over bands to the two flat bands of a single spin direction and valley $\tau=K$. The filling factor of this flavor with two bands will be denoted by $\nu_{K}\in[-1,1]$. The magnetization vs. filling is plotted for a typical magnitude of strain in Fig.~\ref{fig:mag_vs_filling_nonint} for five different twist angles, $\theta\in\lbrace0.95^{\circ},1.07^{\circ},1.1^{\circ},1.3^{\circ},1.5^{\circ}\rbrace$. With our choice of model parameters, the first magic angle is at $\theta_{MA}\approx 1.07^{\circ}$, and the flat bands are separated by a gap from the remote bands for $\theta \gtrsim 0.9^{\circ}$. Fig.~\ref{fig:mag_vs_filling_nonint} shows three key points: $(i)$ It is possible to obtain a significant in-plane orbital magnetization in the non-interacting model using strain, provided that flavor symmetry is broken, $(ii)$ the magnetization is highly sensitive to the twist angle, peaking in the vicinity of the magic angle, and $(iii)$ the conduction band exactly cancels the valence band contribution. This final point implies that filled flavors do not contribute to an overall magnetization, and we find that a significant response occurs near half filling of a single flavor.

\begin{figure}[t]
    \centering
    \includegraphics[width=0.5\textwidth]{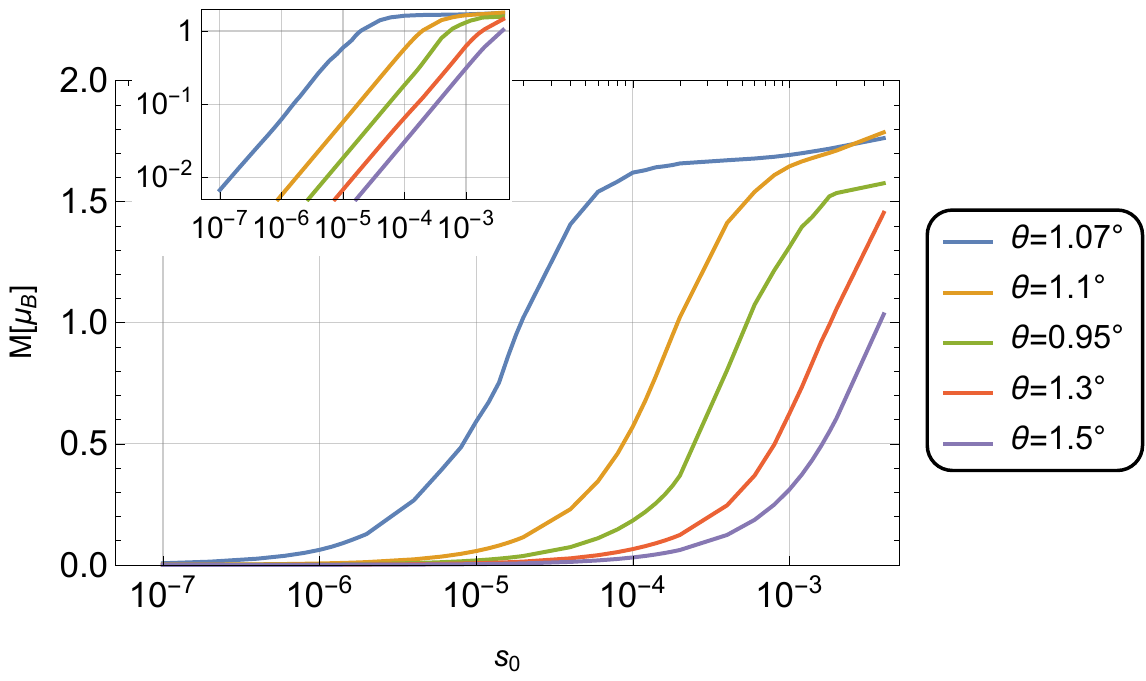}
    \caption{The valley-$K$ resolved magnitude of the magnetization \textit{vs} strain at $\nu_{K}=0$ for the noninteracting TBG. The horizontal axis $s_{0}$ denotes the  principal strains ($s'=-s''=s_{0}$) of the heterostrain with $\phi_{S}=45^{\circ}$, and is bounded by $s_{0}/\theta\lesssim0.1$ such that the system is dominantly ``twisted". Note that the strain necessary to obtain a significant magnetization is extremely small near the magic angle. Inset presents the data on a log-log plot, and shows the linear dependence at small strain.}
    \label{fig:mag_vs_strain_nonint}
\end{figure}

Near the magic angle, the magnetization peaks when completely filling the valence band ($\nu_{K}=0$). Therefore, we plot the magnetization vs. strain at this filling in Fig.~\ref{fig:mag_vs_strain_nonint} for several twist angles. The key result from this figure is that, near the magic angle, the strain magnitude necessary to reach $M_{\parallel}\sim1\mu_{B}$ is extremely small, even in comparison with the twist angle, which sets a reference scale of $\theta\sim0.02$. For small strain one observes (inset) the anticipated linear dependence on strain magnitude. The linear response is associated with a magnetization magnitude that is independent of the strain direction (Appendix \ref{sec:app_symmetry}). At higher strain values, the magnitude of the magnetization becomes generally dependent on strain direction. Extreme strains approaching the twist angle are beyond the scope of this paper. In particular, for $s'=-s''=s_{0}$ and $\phi_{S}=45^{\circ}$, the moir\'{e} pattern is a square lattice for $s_{0}/\theta=2-\sqrt{3}\approx 0.268$. This motivates us to consider $s_{0}/\theta\lesssim0.1$ as a bound for the strain under which the system is dominantly ``twisted".

\subsection{Magnetization of valley polarized states in the presence of interactions}

We now turn our attention to the effect of electron-electron interactions. As previously discussed, $C_3$ symmetry breaking is insufficient by itself for producing a net orbital magnetization - one must also break $C_2$ and $\mathcal{T}$ symmetries. In MATBG, these two symmetries may be spontaneously broken due to electron-electron interactions, which possibly lead to ground states that are valley polarized, spin polarized or both.
In this section, we use a simplified Hartree-Fock procedure to calculate the orbital magnetization of valley polarized states, including effects of band renormalizations due to interactions.

\begin{table}[]
    \centering
    \begin{tabular}{|c||c|c|c|c||c|c|c|c|}
        \hline
        & \multicolumn{4}{c||}{Initial: $\overrightarrow{\nu_{i}}$} & \multicolumn{4}{c|}{Final: $\overrightarrow{\nu_{f}}$}\\
        \hline
        Configuration
        & $\nu_{K\uparrow}^{i}$ & $\nu_{K'\downarrow}^{i}$ & $\nu_{K\downarrow}^{i}$ & $\nu_{K'\uparrow}^{i}$ 
        & $\nu_{K\uparrow}^{f}$& $\nu_{K'\downarrow}^{f}$ & $\nu_{K\downarrow}^{f}$ & $\nu_{K'\uparrow}^{f}$ \\
        \hline
        \hline
        Unpolarized & $-1$ & $-1$ & $-1$ & $-1$ & $+1$ & $+1$ & $+1$ & $+1$\\
        \hline
        VP $\nu=+3$ & $-1$ & $+1$ & $+1$ & $+1$ & $+1$ & $+1$ & $+1$ & $+1$\\
        \hline
        VP $\nu=+1$ & $-1$ & $+1$ & $-1$ & $-1$ & $+1$ & $+1$ & $+1$ & $+1$\\
        \hline
        VP $\nu=-1$ & $-1$ & $-1$ & $-1$ & $-1$ & $+1$ & $-1$ & $+1$ & $+1$\\
        \hline
        VP $\nu=-3$ & $-1$ & $-1$ & $-1$ & $-1$ & $+1$ & $-1$ & $-1$ & $-1$\\
        \hline
    \end{tabular}
    \caption{Flavor interpolation scheme for unpolarized and valley-polarized (VP) states. A completely empty (filled) flavor $\lbrace\tau,s\rbrace$ is represented by a filling factor of $\nu_{\tau,s}^{i/f}=-1$ ($\nu_{\tau,s}^{i/f}=+1$). The density matrix is constructed for the two limits labeled here by ``Initial" and ``Final" using the non-interacting states, and then interpolated for intermediate fillings. The VP states with $\nu\in\lbrace-3,-1,+1,+3\rbrace$ are obtained by the halfway point between the initial and final points defined by $\protect\overrightarrow{\nu_{i}}$ and $\protect\overrightarrow{\nu_{f}}$.}
    \label{tab:flavor_interpolation}
\end{table}

\subsubsection{Calculation method for the ground state energy and magnetization}

The orbital magnetization in Eq.~\eqref{eq:inplane_magnetization_def} requires us to evaluate the gradient of the ground state energy with respect to magnetic field. 
A standard approximation procedure for obtaining the ground state of an interacting Hamiltonian $\hat{H}$ involves a self-consistent Hartree-Fock (SCHF) procedure. In this procedure one minimizes the expectation energy $E=\langle \hat{H}\rangle$ over a family of Slater determinant wavefunctions by self-consistently solving for the ground state of the single-particle Hartree-Fock Hamiltonian:
\begin{equation}\label{eq:schf_def}
    \hat{H}_{HF}=\hat{H}_{0}+\hat{\Sigma}^{HF}[\rho],
\end{equation}
where $\hat{H}_{0}$ is a non-interacting Hamiltonian, $\hat{\Sigma}^{HF}[\rho]$ is the self energy operator in the Hartree-Fock approximation, and $\rho$ is the single-particle density matrix corresponding to the ground state. The explicit form of these operators, including the scheme to subtract the interaction effects of a pair of decoupled layers\cite{Xie2020,Xie2021}, is given in Appendix~\ref{sec:app_hartree_fock}. 

The SCHF procedure performed at filling $\nu$ and magnetic field $\vec{B}_{\parallel}$ provides the chemical potential $\mu(\nu,\vec{B}_{\parallel})$ as the lowest quasiparticle excitation energy. The ground state energy can be obtained by integrating the chemical potential over the filling, and the orbital magnetization in Eq.~\eqref{eq:inplane_magnetization_def} can be calculated by:
\begin{equation}\label{eq:mag_from_chem_pot}
    \vec{M}_{\parallel}(\nu)
    = - \int_{\nu_{0}}^{\nu} \left.\frac{\partial\mu(\nu^{\prime},\vec{B}_{\parallel})}{\partial \vec{B}_{\parallel}}\right|_{\vec{B}_{\parallel}=0}d\nu^{\prime}+\vec{M}_{\parallel}(\nu_{0}).
\end{equation}

The self-consistent approach for determining $\rho(\nu)$ in Eq.~\eqref{eq:schf_def} is computationally expensive. Instead, we employ an interpolation scheme similar to the work by Xie and MacDonald (Ref.~\onlinecite{Xie2021}) for \textit{approximating} the density matrix for arbitrary filling without performing a self-consistent calculation. In this method, we take two single-particle density matrices at two filling configurations, denoted by $\rho(\nu_{i})$ and $\rho(\nu_{f})$, and assume that the density matrix $\rho(\nu)$ at any intermediate filling is a linear interpolation between the two limits. Neglecting any interaction-driven hybridization between remote and flat bands, the many-electron ground state at $\nu=\pm4$ becomes the non-interacting single-particle ground state, thereby providing exact density matrices for $\nu_{i}=-4$ and $\nu_{f}=+4$.

We are interested in valley-polarized (VP) states, which cannot be adequately described by filling all flavors equally. Therefore we consider individual flavor filling components $\ovarrow{\nu}=(\nu_{K\uparrow},\nu_{K'\downarrow},\nu_{K\downarrow},\nu_{K'\uparrow})$, such that the total filling $\nu$ is the sum of the four components of $\ovarrow{\nu}$. We perform a linear interpolation of the density matrix in this four dimensional parameter space between two configuration points $\ovarrow{\nu_{i}}$ and $\ovarrow{\nu_{f}}$: 
\begin{equation}\label{eq:density_interpolation_with_flavors}
    \rho(\nu) = 
    \frac{\nu_{f}-\nu}{\nu_{f}-\nu_{i}}\rho(\ovarrow{\nu_{i}})
    +\frac{\nu-\nu_{i}}{\nu_{f}-\nu_{i}}\rho(\ovarrow{\nu_{f}})
    ,\quad \nu\in[\nu_{i},\nu_{f}],
\end{equation}
where $\rho(\ovarrow{\nu_{i}})$ and $\rho(\ovarrow{\nu_{f}})$ are diagonal in flavor, with an overall filling factor $\nu_{i}$ and $\nu_{f}$, respectively.
If the individual flavors in $\ovarrow{\nu_{i}}$ (or $\ovarrow{\nu_{f}}$)  are either empty or full, each flavor block in the density matrix $\rho(\ovarrow{\nu_{i}})$ (or $\rho(\ovarrow{\nu_{f}})$) is given by the corresponding density matrix of the non-interacting BM model.

The choice of initial and final density matrices for interpolation determines the nature of the ground state under consideration. Consider the VP state at $\nu=+1$ which is characterized by $\ovarrow{\nu}=(0,+1,0,0)$, and consists of one filled flavor and three half filled flavors. 
Within our scheme, the density matrix of such a state can be obtained by interpolating between a $\nu_{i}=-2$ initial state in which three flavors are empty and one is full,  $\ovarrow{\nu_{i}}=(-1,+1,-1,-1)$, and a $\nu_{f}=+4$ final state in which all flavors are full, $\ovarrow{\nu_{f}}=(+1,+1,+1,+1)$. The interpolation scheme used to describe other valley-polarized states is summarized in Table.~\ref{tab:flavor_interpolation}.

The calculation method for the magnetization is given by the following steps. First, one chooses a VP state (e.g. $\nu=+1$). Next, the initial and final configuration points for the interpolarion are selected from Table.~\ref{tab:flavor_interpolation}, and the density matrix is calculated for every filling $\nu\in[\nu_{i},\nu_{f}]$ using Eq.~\eqref{eq:density_interpolation_with_flavors}. Given the interpolated density matrix $\rho(\nu)$, we compute the Hartree-Fock Hamiltonian in Eq.~\eqref{eq:schf_def} without self-consistency. The quasiparticle spectrum of this Hartree-Fock Hamiltonian provides our approximate chemical potential $\mu(\nu)$, which is then integrated to find the ground state energy. Repeating the procedure for different magnetic fields allows one to compute the orbital magnetization using Eq.~\eqref{eq:mag_from_chem_pot}. Finally, the integration constant is set to nullify the magnetization when all interpolated flavors are completely empty or filled, which in Table~\ref{tab:flavor_interpolation} is always the case for either $\nu=\nu_{i}$ or $\nu=\nu_{f}$. Numerical parameters and further technical details are given in Appendix~\ref{sec:app_numerical}.

In every interpolation configuration in Table~\ref{tab:flavor_interpolation} we set one pair of flavors related by $\mathcal{T}$ to have equal filling at both initial and final points, setting $\nu_{K\downarrow}(\nu)=\nu_{K'\uparrow}(\nu)$ at every intermediate filling $\nu$. By construction, these flavor's contributions to the magnetization cancel each other. A third flavor filling $\nu_{K'\downarrow}$ is kept fixed throughout the interpolation, and therefore can only contribute to the magnetization's integration constant. Therefore, for our purpose of calculating the orbital magnetization, this scheme allows us to focus on a single flavor block in the Hartree-Fock Hamiltonian, with points of subtlety discussed in Appendix~\ref{sec:app_single_flavor}. 

\subsubsection{Magnetization results}

\begin{figure*}
    \centering
    \includegraphics[width=7.0in]{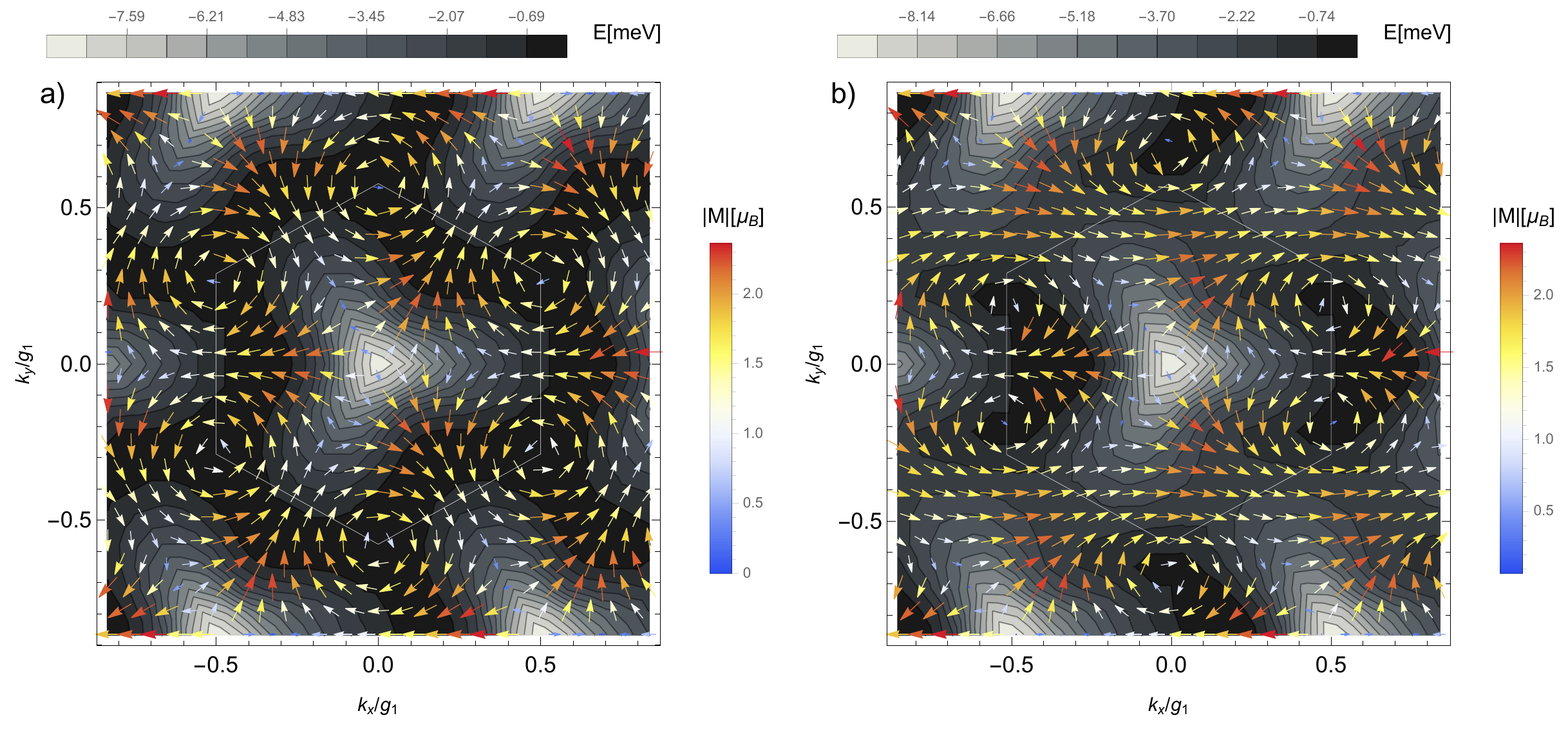}
    \caption{Orbital in-plane magnetization vectors of individual Hartree-Fock eigenstates at filling $\nu=+1$ in the valence band of valley $\tau=K$, plotted across the moir\'{e} BZ for the VP $\nu=+1$ state of TBG with twist angle of $\theta=1.1^{\circ}$. In this state, $\nu_{K'\downarrow}=+1$ and all other flavors are at zero filling. Background contours represent the quasiparticle energy band. a) No strain, b) Heterostrain as specified in Fig.~\ref{fig:mag_vs_bz_nonint}b. Individual magnetization vectors are similar in magnitude to those obtained in the absence of interactions (Fig.~\ref{fig:mag_vs_bz_nonint}), but their orientations are somewhat less sensitive to heterostrain once interactions are introduced. The hexagon outlines the moir\'{e} BZ.
    }
    \label{fig:mag_vs_bz_int}
\end{figure*}
\begin{figure}[t]
    \centering
    \includegraphics[width=0.5\textwidth]{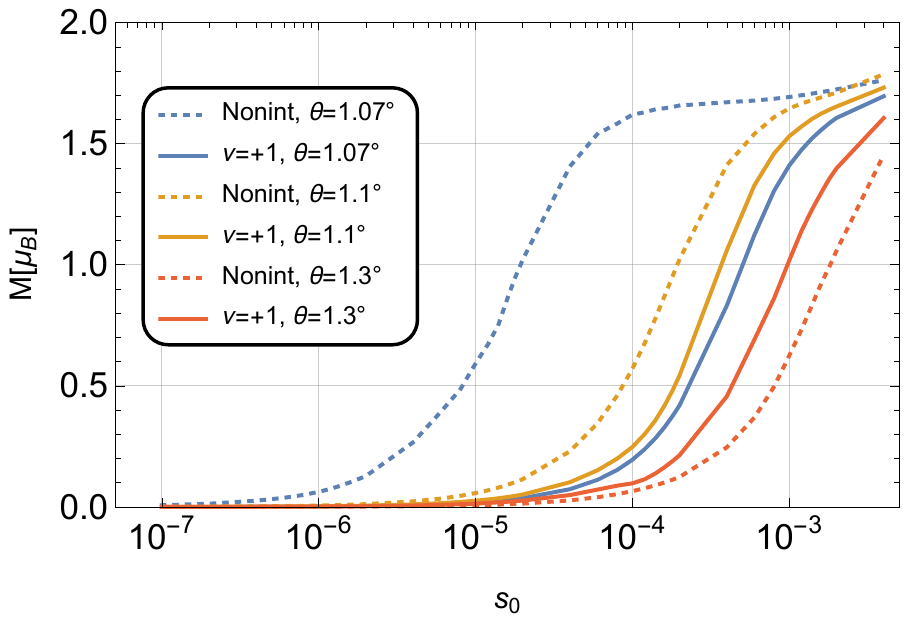}
    \caption{The in-plane orbital magnetization magnitude per moir\'{e} unit cell \textit{vs} strain, for $\nu=+1$ VP states at differents twist angle, compared with their non-interacting counterparts. The horizontal axis $s_{0}$ denotes the principal strains as described in Fig.~\ref{fig:mag_vs_strain_nonint}. Note that the plots of the non-interacting model are valley-$K$ resolved, while the VP states display a net magnetization. We find that in the vicinity of the magic angle the interactions reduce sensitivity to strain, yet significant magnetization is still obtained at small strains.}
    \label{fig:mag_vs_strain_int}
\end{figure}

Fig.~\ref{fig:mag_vs_bz_int} compares between the magnetizations of individual Hartree-Fock eigenstates of the unstrained and strained TBGs at filling $\nu=+1$, across the moir\'{e} BZ. The calculation is done for the VP $\nu=+1$ configuration and presented for the valence band of valley $K$.
We find that, similarly to the non-interacting case, individual states contribute $\sim 1\mu_{B}$ with and without strain, but the 2D orientations seem more resilient to heterostrain in the presence of interactions. In a sense, it seems that interactions stabilize the $C_3$ symmetry.

In the presence of interactions, the total magnetization of the TBG is not the sum of the magnetizations of the individual states plotted in Fig.~\ref{fig:mag_vs_bz_int}. The reason is that the individual Hartree-Fock eigenvalues depend on filling, which is varied with each additional particle. Thus, the correct way to calculate the magnetization is by means of Eq.~\eqref{eq:mag_from_chem_pot}.

In Fig.~\ref{fig:mag_vs_strain_int} we plot a comparison between the magnetizations of $\nu=+1$ VP states at different twist angles and their non-interacting counterparts, as a function of strain. We find that, in the vicinity of the magic angle, interactions make the magnetization less sensitive to strain. Furthermore, the dependencies of the magnetizations on strain for the three twist angles we examined become closer to each other. However, note that even when interactions are taken into account, we obtain a significant orbital magnetization (larger than $1\mu_B$ per electron), even for strains as small as $s_0\approx 4\cdot 10^{-4}$. We comment that the $\nu=+3$ VP states behave similarly to the $\nu=+1$ VP states. Due to our approximate particle-hole symmetry, the negative fillings states $\nu=-1$ and $\nu=-3$ show the same magnetization magnitude as their symmetric counterparts.

We find an empirical correspondence between the bandwidth of the individual (quasi-)particle spectrum and the strength of the magnetization response, regardless of whether the spectrum originated from the interacting or non-interacting model (Appendix~\ref{sec:app_hartree_fock}). This could be interpreted as a consequence of the change in the Dirac velocity caused by interactions, which mimics a change in the twist angle.

\section{Discussion and conclusions}\label{sec:discussion}

In this paper we have shown that valley-polarized states in heterostrained MATBG have a significant in-plane orbital magnetization, which may approach the order of $\sim 1\mu_{B}$ per moir\'{e} unit cell. This orbital magnetization can be thought of as arising from individual magnetic moments of Bloch states partially aligning due to $C_{3}$ symmetry breaking, with their time-reversed partners unoccupied due to the state being valley-polarized. The flatness of the bands in the vicinity of the magic angle plays a dual role in this picture: it increases the relative strength of electron-electron interaction, which allows for the $C_{2}$ and $\mathcal{T}$-broken valley polarized states, and it increases the sensitivity of the band structure to $C_{3}$ symmetry breaking due to heterostrain. Band renormalization due to interactions is accounted for using a simplified model, and near the magic angle its main effect is an increase in bandwidth and consequently a slight decrease in sensitivity to strain. Even in the presence of this band renormalization, typical strain values of $s_{0}\sim 10^{-4}-10^{-3}$ \cite{Kerelsky2019,Choi2019,Xie2019,Kazmierczak} are enough to generate an orbital magnetization of $\sim 1\mu_{B}$ per moir\'{e} unit cell for twist angles near the magic angle, a surprising result considering that $s_{0}/\theta \sim 0.01-0.1$ in this parameter regime.

Our conclusions are of direct relevance to experiments which exhibit evidence of flavor symmetry breaking in the form of spin and valley polarized states at $\nu=+1,+3$\cite{Zondiner2020,Park2021,Sharpe2019,Serlin2020}. In these states, the total magnetic moment comprises of orbital and spin components, with the latter being at most $1\mu_{B}$ per moir\'{e} unit cell. Any magnetization in excess of this value is a strong indication of an orbital component. There is a further signature that our model predicts - the orbital component is determined by the direction of the strain, as opposed to the isotropic spin magnetization.

In fact, it is possible that this orbital component has already been observed in Ref.~\onlinecite{Zondiner2020}, where the in-plane magnetization was extracted from electronic compressibility measurements in the presence of a field. Close to the magic angle and near $\nu=1$, the in-plane magnetization exceeds $1\mu_B$ per unit cell. As expected, this effect disappears away from the magic angle. However, the direction of the magnetization in the plane was not studied in the experiment. Our analysis predicts the relation of this direction to the strain.  

Our result suggests that in the presence of strain, an in-plane magnetic field induces a valley-Zeeman effect, which favours valley-polarized states. In a sample that contains domains with different strains, we expect the in-plane magnetization to be different in each domain, resulting in an out-of-plane field at the domain boundaries. We emphasize that in our model we assumed that our system has $C_{2}\mathcal{T}$ symmetry (there is no sublattice potential), and hence there is no out-of-plane magnetization in the bulk, in contrast with the system described in Refs.~\onlinecite{Zhang2020a,He2020,Bultinck2020}. 

For future works, we suggest exploring the in-plane magnetization of nematic states where the $C_{3}$ symmetry is broken spontaneously~\cite{Liu2021nematic,Cao2021,Choi2019,Jiang2019}. 
In such a phase, the direction of the magnetization is presumably related to the orientation of the nematic order parameter. Therefore, one could imagine training the nematic order using an in-plane magnetic field. 

\acknowledgements
We thank Keshav Pareek and Achim Rosch for useful discussions. We acknowledge support from the Israeli Science Foundation Quantum Science and Technology grant no. 2074/19, and from the
CRC 183 of the Deutsche Forschungsgemeinschaft (Project C02). 
This work has received funding from the European Research Council (ERC) under the European Union’s Horizon 2020 research and innovation programme [grant agreements No. 788715 and 817799, Projects LEGOTOP (AS) and HQMAT (EB)].

\bibliography{strained_tbg_arxiv.bib}

\begin{thebibliography}{42}%
\makeatletter
\providecommand \@ifxundefined [1]{%
 \@ifx{#1\undefined}
}%
\providecommand \@ifnum [1]{%
 \ifnum #1\expandafter \@firstoftwo
 \else \expandafter \@secondoftwo
 \fi
}%
\providecommand \@ifx [1]{%
 \ifx #1\expandafter \@firstoftwo
 \else \expandafter \@secondoftwo
 \fi
}%
\providecommand \natexlab [1]{#1}%
\providecommand \enquote  [1]{``#1''}%
\providecommand \bibnamefont  [1]{#1}%
\providecommand \bibfnamefont [1]{#1}%
\providecommand \citenamefont [1]{#1}%
\providecommand \href@noop [0]{\@secondoftwo}%
\providecommand \href [0]{\begingroup \@sanitize@url \@href}%
\providecommand \@href[1]{\@@startlink{#1}\@@href}%
\providecommand \@@href[1]{\endgroup#1\@@endlink}%
\providecommand \@sanitize@url [0]{\catcode `\\12\catcode `\$12\catcode
  `\&12\catcode `\#12\catcode `\^12\catcode `\_12\catcode `\%12\relax}%
\providecommand \@@startlink[1]{}%
\providecommand \@@endlink[0]{}%
\providecommand \url  [0]{\begingroup\@sanitize@url \@url }%
\providecommand \@url [1]{\endgroup\@href {#1}{\urlprefix }}%
\providecommand \urlprefix  [0]{URL }%
\providecommand \Eprint [0]{\href }%
\providecommand \doibase [0]{http://dx.doi.org/}%
\providecommand \selectlanguage [0]{\@gobble}%
\providecommand \bibinfo  [0]{\@secondoftwo}%
\providecommand \bibfield  [0]{\@secondoftwo}%
\providecommand \translation [1]{[#1]}%
\providecommand \BibitemOpen [0]{}%
\providecommand \bibitemStop [0]{}%
\providecommand \bibitemNoStop [0]{.\EOS\space}%
\providecommand \EOS [0]{\spacefactor3000\relax}%
\providecommand \BibitemShut  [1]{\csname bibitem#1\endcsname}%
\let\auto@bib@innerbib\@empty
\bibitem [{\citenamefont {Bistritzer}\ and\ \citenamefont
  {MacDonald}(2011)}]{Bistritzer2011}%
  \BibitemOpen
  \bibfield  {author} {\bibinfo {author} {\bibfnamefont {Rafi}\ \bibnamefont
  {Bistritzer}}\ and\ \bibinfo {author} {\bibfnamefont {Allan~H.}\ \bibnamefont
  {MacDonald}},\ }\bibfield  {title} {\enquote {\bibinfo {title} {{Moir{\'{e}}
  bands in twisted double-layer graphene}},}\ }\href {\doibase
  10.1073/pnas.1108174108} {\bibfield  {journal} {\bibinfo  {journal}
  {Proceedings of the National Academy of Sciences of the United States of
  America}\ }\textbf {\bibinfo {volume} {108}},\ \bibinfo {pages}
  {12233--12237} (\bibinfo {year} {2011})},\ \Eprint
  {http://arxiv.org/abs/1009.4203} {arXiv:1009.4203} \BibitemShut {NoStop}%
\bibitem [{\citenamefont {Cao}\ \emph {et~al.}(2018{\natexlab{a}})\citenamefont
  {Cao}, \citenamefont {Fatemi}, \citenamefont {Demir}, \citenamefont {Fang},
  \citenamefont {Tomarken}, \citenamefont {Luo}, \citenamefont
  {Sanchez-Yamagishi}, \citenamefont {Watanabe}, \citenamefont {Taniguchi},
  \citenamefont {Kaxiras}, \citenamefont {Ashoori},\ and\ \citenamefont
  {Jarillo-Herrero}}]{Cao2018}%
  \BibitemOpen
  \bibfield  {author} {\bibinfo {author} {\bibfnamefont {Yuan}\ \bibnamefont
  {Cao}}, \bibinfo {author} {\bibfnamefont {Valla}\ \bibnamefont {Fatemi}},
  \bibinfo {author} {\bibfnamefont {Ahmet}\ \bibnamefont {Demir}}, \bibinfo
  {author} {\bibfnamefont {Shiang}\ \bibnamefont {Fang}}, \bibinfo {author}
  {\bibfnamefont {Spencer~L.}\ \bibnamefont {Tomarken}}, \bibinfo {author}
  {\bibfnamefont {Jason~Y.}\ \bibnamefont {Luo}}, \bibinfo {author}
  {\bibfnamefont {Javier~D.}\ \bibnamefont {Sanchez-Yamagishi}}, \bibinfo
  {author} {\bibfnamefont {Kenji}\ \bibnamefont {Watanabe}}, \bibinfo {author}
  {\bibfnamefont {Takashi}\ \bibnamefont {Taniguchi}}, \bibinfo {author}
  {\bibfnamefont {Efthimios}\ \bibnamefont {Kaxiras}}, \bibinfo {author}
  {\bibfnamefont {Ray~C.}\ \bibnamefont {Ashoori}}, \ and\ \bibinfo {author}
  {\bibfnamefont {Pablo}\ \bibnamefont {Jarillo-Herrero}},\ }\bibfield  {title}
  {\enquote {\bibinfo {title} {{Correlated insulator behaviour at half-filling
  in magic-angle graphene superlattices}},}\ }\href {\doibase
  10.1038/nature26154} {\bibfield  {journal} {\bibinfo  {journal} {Nature}\
  }\textbf {\bibinfo {volume} {556}},\ \bibinfo {pages} {80--84} (\bibinfo
  {year} {2018}{\natexlab{a}})}\BibitemShut {NoStop}%
\bibitem [{\citenamefont {Cao}\ \emph {et~al.}(2018{\natexlab{b}})\citenamefont
  {Cao}, \citenamefont {Fatemi}, \citenamefont {Fang}, \citenamefont
  {Watanabe}, \citenamefont {Taniguchi}, \citenamefont {Kaxiras},\ and\
  \citenamefont {Jarillo-Herrero}}]{Cao2018a}%
  \BibitemOpen
  \bibfield  {author} {\bibinfo {author} {\bibfnamefont {Yuan}\ \bibnamefont
  {Cao}}, \bibinfo {author} {\bibfnamefont {Valla}\ \bibnamefont {Fatemi}},
  \bibinfo {author} {\bibfnamefont {Shiang}\ \bibnamefont {Fang}}, \bibinfo
  {author} {\bibfnamefont {Kenji}\ \bibnamefont {Watanabe}}, \bibinfo {author}
  {\bibfnamefont {Takashi}\ \bibnamefont {Taniguchi}}, \bibinfo {author}
  {\bibfnamefont {Efthimios}\ \bibnamefont {Kaxiras}}, \ and\ \bibinfo {author}
  {\bibfnamefont {Pablo}\ \bibnamefont {Jarillo-Herrero}},\ }\bibfield  {title}
  {\enquote {\bibinfo {title} {{Unconventional superconductivity in magic-angle
  graphene superlattices}},}\ }\href {\doibase 10.1038/nature26160} {\bibfield
  {journal} {\bibinfo  {journal} {Nature}\ }\textbf {\bibinfo {volume} {556}},\
  \bibinfo {pages} {43--50} (\bibinfo {year} {2018}{\natexlab{b}})}\BibitemShut
  {NoStop}%
\bibitem [{\citenamefont {Yankowitz}\ \emph {et~al.}(2019)\citenamefont
  {Yankowitz}, \citenamefont {Chen}, \citenamefont {Polshyn}, \citenamefont
  {Zhang}, \citenamefont {Watanabe}, \citenamefont {Taniguchi}, \citenamefont
  {Graf}, \citenamefont {Young},\ and\ \citenamefont {Dean}}]{Yankowitz2019}%
  \BibitemOpen
  \bibfield  {author} {\bibinfo {author} {\bibfnamefont {Matthew}\ \bibnamefont
  {Yankowitz}}, \bibinfo {author} {\bibfnamefont {Shaowen}\ \bibnamefont
  {Chen}}, \bibinfo {author} {\bibfnamefont {Hryhoriy}\ \bibnamefont
  {Polshyn}}, \bibinfo {author} {\bibfnamefont {Yuxuan}\ \bibnamefont {Zhang}},
  \bibinfo {author} {\bibfnamefont {K}~\bibnamefont {Watanabe}}, \bibinfo
  {author} {\bibfnamefont {T}~\bibnamefont {Taniguchi}}, \bibinfo {author}
  {\bibfnamefont {David}\ \bibnamefont {Graf}}, \bibinfo {author}
  {\bibfnamefont {Andrea~F}\ \bibnamefont {Young}}, \ and\ \bibinfo {author}
  {\bibfnamefont {Cory~R}\ \bibnamefont {Dean}},\ }\bibfield  {title} {\enquote
  {\bibinfo {title} {{Tuning superconductivity in twisted bilayer graphene}},}\
  }\href {http://science.sciencemag.org/} {\bibfield  {journal} {\bibinfo
  {journal} {Science}\ } (\bibinfo {year} {2019})}\BibitemShut {NoStop}%
\bibitem [{\citenamefont {Arora}\ \emph {et~al.}(2020)\citenamefont {Arora},
  \citenamefont {Polski}, \citenamefont {Zhang}, \citenamefont {Thomson},
  \citenamefont {Choi}, \citenamefont {Kim}, \citenamefont {Lin}, \citenamefont
  {Wilson}, \citenamefont {Xu}, \citenamefont {Chu}, \citenamefont {Watanabe},
  \citenamefont {Taniguchi}, \citenamefont {Alicea}, \citenamefont
  {Nadj-Perge},\ and\ \citenamefont {Zhang}}]{Arora2020}%
  \BibitemOpen
  \bibfield  {author} {\bibinfo {author} {\bibfnamefont {Harpreet~Singh}\
  \bibnamefont {Arora}}, \bibinfo {author} {\bibfnamefont {Robert}\
  \bibnamefont {Polski}}, \bibinfo {author} {\bibfnamefont {Yiran}\
  \bibnamefont {Zhang}}, \bibinfo {author} {\bibfnamefont {Alex}\ \bibnamefont
  {Thomson}}, \bibinfo {author} {\bibfnamefont {Youngjoon}\ \bibnamefont
  {Choi}}, \bibinfo {author} {\bibfnamefont {Hyunjin}\ \bibnamefont {Kim}},
  \bibinfo {author} {\bibfnamefont {Zhong}\ \bibnamefont {Lin}}, \bibinfo
  {author} {\bibfnamefont {Ilham~Zaky}\ \bibnamefont {Wilson}}, \bibinfo
  {author} {\bibfnamefont {Xiaodong}\ \bibnamefont {Xu}}, \bibinfo {author}
  {\bibfnamefont {Jiun-Haw}\ \bibnamefont {Chu}}, \bibinfo {author}
  {\bibfnamefont {Kenji}\ \bibnamefont {Watanabe}}, \bibinfo {author}
  {\bibfnamefont {Takashi}\ \bibnamefont {Taniguchi}}, \bibinfo {author}
  {\bibfnamefont {Jason}\ \bibnamefont {Alicea}}, \bibinfo {author}
  {\bibfnamefont {Stevan}\ \bibnamefont {Nadj-Perge}}, \ and\ \bibinfo {author}
  {\bibfnamefont {Yiran}\ \bibnamefont {Zhang}},\ }\bibfield  {title} {\enquote
  {\bibinfo {title} {{Superconductivity in metallic twisted bilayer graphene
  stabilized by WSe 2}},}\ }\href {\doibase 10.1038/s41586-020-2473-8}
  {\bibfield  {journal} {\bibinfo  {journal} {Nature}\ }\textbf {\bibinfo
  {volume} {583}},\ \bibinfo {pages} {379} (\bibinfo {year}
  {2020})}\BibitemShut {NoStop}%
\bibitem [{\citenamefont {Choi}\ \emph {et~al.}(2019)\citenamefont {Choi},
  \citenamefont {Kemmer}, \citenamefont {Peng}, \citenamefont {Thomson},
  \citenamefont {Arora}, \citenamefont {Polski}, \citenamefont {Zhang},
  \citenamefont {Ren}, \citenamefont {Alicea}, \citenamefont {Refael},
  \citenamefont {von Oppen}, \citenamefont {Watanabe}, \citenamefont
  {Taniguchi},\ and\ \citenamefont {Nadj-Perge}}]{Choi2019}%
  \BibitemOpen
  \bibfield  {author} {\bibinfo {author} {\bibfnamefont {Youngjoon}\
  \bibnamefont {Choi}}, \bibinfo {author} {\bibfnamefont {Jeannette}\
  \bibnamefont {Kemmer}}, \bibinfo {author} {\bibfnamefont {Yang}\ \bibnamefont
  {Peng}}, \bibinfo {author} {\bibfnamefont {Alex}\ \bibnamefont {Thomson}},
  \bibinfo {author} {\bibfnamefont {Harpreet}\ \bibnamefont {Arora}}, \bibinfo
  {author} {\bibfnamefont {Robert}\ \bibnamefont {Polski}}, \bibinfo {author}
  {\bibfnamefont {Yiran}\ \bibnamefont {Zhang}}, \bibinfo {author}
  {\bibfnamefont {Hechen}\ \bibnamefont {Ren}}, \bibinfo {author}
  {\bibfnamefont {Jason}\ \bibnamefont {Alicea}}, \bibinfo {author}
  {\bibfnamefont {Gil}\ \bibnamefont {Refael}}, \bibinfo {author}
  {\bibfnamefont {Felix}\ \bibnamefont {von Oppen}}, \bibinfo {author}
  {\bibfnamefont {Kenji}\ \bibnamefont {Watanabe}}, \bibinfo {author}
  {\bibfnamefont {Takashi}\ \bibnamefont {Taniguchi}}, \ and\ \bibinfo {author}
  {\bibfnamefont {Stevan}\ \bibnamefont {Nadj-Perge}},\ }\bibfield  {title}
  {\enquote {\bibinfo {title} {{Electronic correlations in twisted bilayer
  graphene near the magic angle}},}\ }\href {\doibase
  10.1038/s41567-019-0606-5} {\bibfield  {journal} {\bibinfo  {journal} {Nature
  Physics}\ }\textbf {\bibinfo {volume} {15}},\ \bibinfo {pages} {1174--1180}
  (\bibinfo {year} {2019})},\ \Eprint {http://arxiv.org/abs/1901.02997}
  {arXiv:1901.02997} \BibitemShut {NoStop}%
\bibitem [{\citenamefont {Wong}\ \emph {et~al.}(2020)\citenamefont {Wong},
  \citenamefont {Nuckolls}, \citenamefont {Oh}, \citenamefont {Lian},
  \citenamefont {Xie}, \citenamefont {Jeon}, \citenamefont {Watanabe},
  \citenamefont {Taniguchi}, \citenamefont {Bernevig},\ and\ \citenamefont
  {Yazdani}}]{Wong2020}%
  \BibitemOpen
  \bibfield  {author} {\bibinfo {author} {\bibfnamefont {Dillon}\ \bibnamefont
  {Wong}}, \bibinfo {author} {\bibfnamefont {Kevin~P}\ \bibnamefont
  {Nuckolls}}, \bibinfo {author} {\bibfnamefont {Myungchul}\ \bibnamefont
  {Oh}}, \bibinfo {author} {\bibfnamefont {Biao}\ \bibnamefont {Lian}},
  \bibinfo {author} {\bibfnamefont {Yonglong}\ \bibnamefont {Xie}}, \bibinfo
  {author} {\bibfnamefont {Sangjun}\ \bibnamefont {Jeon}}, \bibinfo {author}
  {\bibfnamefont {Kenji}\ \bibnamefont {Watanabe}}, \bibinfo {author}
  {\bibfnamefont {Takashi}\ \bibnamefont {Taniguchi}}, \bibinfo {author}
  {\bibfnamefont {B~Andrei}\ \bibnamefont {Bernevig}}, \ and\ \bibinfo {author}
  {\bibfnamefont {Ali}\ \bibnamefont {Yazdani}},\ }\bibfield  {title} {\enquote
  {\bibinfo {title} {{Cascade of electronic transitions in magic-angle twisted
  bilayer graphene Sequence of spectroscopic transitions}},}\ }\href {\doibase
  10.1038/s41586-020-2339-0} {\bibfield  {journal} {\bibinfo  {journal}
  {Nature}\ }\textbf {\bibinfo {volume} {582}} (\bibinfo {year} {2020}),\
  10.1038/s41586-020-2339-0}\BibitemShut {NoStop}%
\bibitem [{\citenamefont {Saito}\ \emph {et~al.}(2020)\citenamefont {Saito},
  \citenamefont {Ge}, \citenamefont {Watanabe}, \citenamefont {Taniguchi},\
  and\ \citenamefont {Young}}]{Saito}%
  \BibitemOpen
  \bibfield  {author} {\bibinfo {author} {\bibfnamefont {Yu}~\bibnamefont
  {Saito}}, \bibinfo {author} {\bibfnamefont {Jingyuan}\ \bibnamefont {Ge}},
  \bibinfo {author} {\bibfnamefont {Kenji}\ \bibnamefont {Watanabe}}, \bibinfo
  {author} {\bibfnamefont {Takashi}\ \bibnamefont {Taniguchi}}, \ and\ \bibinfo
  {author} {\bibfnamefont {Andrea~F}\ \bibnamefont {Young}},\ }\bibfield
  {title} {\enquote {\bibinfo {title} {{Independent superconductors and
  correlated insulators in twisted bilayer graphene}},}\ }\href {\doibase
  10.1038/s41567-020-0928-3} {\bibfield  {journal} {\bibinfo  {journal} {Nature
  Physics}\ } (\bibinfo {year} {2020}),\ 10.1038/s41567-020-0928-3}\BibitemShut
  {NoStop}%
\bibitem [{\citenamefont {Stepanov}\ \emph {et~al.}(2020)\citenamefont
  {Stepanov}, \citenamefont {Das}, \citenamefont {Lu}, \citenamefont
  {Fahimniya}, \citenamefont {Watanabe}, \citenamefont {Taniguchi},
  \citenamefont {Koppens}, \citenamefont {Lischner}, \citenamefont {Levitov},\
  and\ \citenamefont {Efetov}}]{Stepanov2020}%
  \BibitemOpen
  \bibfield  {author} {\bibinfo {author} {\bibfnamefont {Petr}\ \bibnamefont
  {Stepanov}}, \bibinfo {author} {\bibfnamefont {Ipsita}\ \bibnamefont {Das}},
  \bibinfo {author} {\bibfnamefont {Xiaobo}\ \bibnamefont {Lu}}, \bibinfo
  {author} {\bibfnamefont {Ali}\ \bibnamefont {Fahimniya}}, \bibinfo {author}
  {\bibfnamefont {Kenji}\ \bibnamefont {Watanabe}}, \bibinfo {author}
  {\bibfnamefont {Takashi}\ \bibnamefont {Taniguchi}}, \bibinfo {author}
  {\bibfnamefont {Frank H~L}\ \bibnamefont {Koppens}}, \bibinfo {author}
  {\bibfnamefont {Johannes}\ \bibnamefont {Lischner}}, \bibinfo {author}
  {\bibfnamefont {Leonid}\ \bibnamefont {Levitov}}, \ and\ \bibinfo {author}
  {\bibfnamefont {Dmitri~K}\ \bibnamefont {Efetov}},\ }\bibfield  {title}
  {\enquote {\bibinfo {title} {{Untying the insulating and superconducting
  orders in magic-angle graphene}},}\ }\href {\doibase
  10.1038/s41586-020-2459-6} {\bibfield  {journal} {\bibinfo  {journal}
  {Nature}\ }\textbf {\bibinfo {volume} {583}},\ \bibinfo {pages} {375}
  (\bibinfo {year} {2020})}\BibitemShut {NoStop}%
\bibitem [{\citenamefont {Uri}\ \emph {et~al.}(2020)\citenamefont {Uri},
  \citenamefont {Grover}, \citenamefont {Cao}, \citenamefont {Crosse},
  \citenamefont {Bagani}, \citenamefont {Rodan-Legrain}, \citenamefont
  {Myasoedov}, \citenamefont {Watanabe}, \citenamefont {Taniguchi},
  \citenamefont {Moon}, \citenamefont {Koshino}, \citenamefont
  {Jarillo-Herrero},\ and\ \citenamefont {Zeldov}}]{Uri2020}%
  \BibitemOpen
  \bibfield  {author} {\bibinfo {author} {\bibfnamefont {A}~\bibnamefont
  {Uri}}, \bibinfo {author} {\bibfnamefont {S}~\bibnamefont {Grover}}, \bibinfo
  {author} {\bibfnamefont {Y}~\bibnamefont {Cao}}, \bibinfo {author}
  {\bibfnamefont {J~A}\ \bibnamefont {Crosse}}, \bibinfo {author}
  {\bibfnamefont {K}~\bibnamefont {Bagani}}, \bibinfo {author} {\bibfnamefont
  {D}~\bibnamefont {Rodan-Legrain}}, \bibinfo {author} {\bibfnamefont
  {Y}~\bibnamefont {Myasoedov}}, \bibinfo {author} {\bibfnamefont
  {K}~\bibnamefont {Watanabe}}, \bibinfo {author} {\bibfnamefont
  {T}~\bibnamefont {Taniguchi}}, \bibinfo {author} {\bibfnamefont
  {P}~\bibnamefont {Moon}}, \bibinfo {author} {\bibfnamefont {M}~\bibnamefont
  {Koshino}}, \bibinfo {author} {\bibfnamefont {P}~\bibnamefont
  {Jarillo-Herrero}}, \ and\ \bibinfo {author} {\bibfnamefont {E}~\bibnamefont
  {Zeldov}},\ }\bibfield  {title} {\enquote {\bibinfo {title} {{Mapping the
  twist-angle disorder and Landau levels in magic-angle graphene}},}\ }\href
  {\doibase 10.1038/s41586-020-2255-3} {\bibfield  {journal} {\bibinfo
  {journal} {Nature}\ }\textbf {\bibinfo {volume} {581}} (\bibinfo {year}
  {2020}),\ 10.1038/s41586-020-2255-3}\BibitemShut {NoStop}%
\bibitem [{\citenamefont {Xie}\ \emph {et~al.}(2019)\citenamefont {Xie},
  \citenamefont {Lian}, \citenamefont {J{\"{a}}ck}, \citenamefont {Liu},
  \citenamefont {Chiu}, \citenamefont {Watanabe}, \citenamefont {Taniguchi},
  \citenamefont {Bernevig},\ and\ \citenamefont {Yazdani}}]{Xie2019}%
  \BibitemOpen
  \bibfield  {author} {\bibinfo {author} {\bibfnamefont {Yonglong}\
  \bibnamefont {Xie}}, \bibinfo {author} {\bibfnamefont {Biao}\ \bibnamefont
  {Lian}}, \bibinfo {author} {\bibfnamefont {Berthold}\ \bibnamefont
  {J{\"{a}}ck}}, \bibinfo {author} {\bibfnamefont {Xiaomeng}\ \bibnamefont
  {Liu}}, \bibinfo {author} {\bibfnamefont {Cheng~Li}\ \bibnamefont {Chiu}},
  \bibinfo {author} {\bibfnamefont {Kenji}\ \bibnamefont {Watanabe}}, \bibinfo
  {author} {\bibfnamefont {Takashi}\ \bibnamefont {Taniguchi}}, \bibinfo
  {author} {\bibfnamefont {B.~Andrei}\ \bibnamefont {Bernevig}}, \ and\
  \bibinfo {author} {\bibfnamefont {Ali}\ \bibnamefont {Yazdani}},\ }\bibfield
  {title} {\enquote {\bibinfo {title} {{Spectroscopic signatures of many-body
  correlations in magic-angle twisted bilayer graphene}},}\ }\href {\doibase
  10.1038/s41586-019-1422-x} {\bibfield  {journal} {\bibinfo  {journal}
  {Nature}\ }\textbf {\bibinfo {volume} {572}},\ \bibinfo {pages} {101--105}
  (\bibinfo {year} {2019})},\ \Eprint {http://arxiv.org/abs/1906.09274}
  {arXiv:1906.09274} \BibitemShut {NoStop}%
\bibitem [{\citenamefont {Rozen}\ \emph {et~al.}(2021)\citenamefont {Rozen},
  \citenamefont {{Min Park}}, \citenamefont {Zondiner}, \citenamefont {Cao},
  \citenamefont {Rodan-Legrain}, \citenamefont {Taniguchi}, \citenamefont
  {Watanabe}, \citenamefont {Oreg}, \citenamefont {Stern}, \citenamefont
  {Berg}, \citenamefont {Jarillo-Herrero},\ and\ \citenamefont
  {Ilani}}]{Rozen2021}%
  \BibitemOpen
  \bibfield  {author} {\bibinfo {author} {\bibfnamefont {Asaf}\ \bibnamefont
  {Rozen}}, \bibinfo {author} {\bibfnamefont {Jeong}\ \bibnamefont {{Min
  Park}}}, \bibinfo {author} {\bibfnamefont {Uri}\ \bibnamefont {Zondiner}},
  \bibinfo {author} {\bibfnamefont {Yuan}\ \bibnamefont {Cao}}, \bibinfo
  {author} {\bibfnamefont {Daniel}\ \bibnamefont {Rodan-Legrain}}, \bibinfo
  {author} {\bibfnamefont {Takashi}\ \bibnamefont {Taniguchi}}, \bibinfo
  {author} {\bibfnamefont {Kenji}\ \bibnamefont {Watanabe}}, \bibinfo {author}
  {\bibfnamefont {Yuval}\ \bibnamefont {Oreg}}, \bibinfo {author}
  {\bibfnamefont {Ady}\ \bibnamefont {Stern}}, \bibinfo {author} {\bibfnamefont
  {Erez}\ \bibnamefont {Berg}}, \bibinfo {author} {\bibfnamefont {Pablo}\
  \bibnamefont {Jarillo-Herrero}}, \ and\ \bibinfo {author} {\bibfnamefont
  {Shahal}\ \bibnamefont {Ilani}},\ }\bibfield  {title} {\enquote {\bibinfo
  {title} {{Entropic evidence for a Pomeranchuk effect in magic-angle
  graphene}},}\ }\href {\doibase 10.1038/s41586-021-03319-3} {\bibfield
  {journal} {\bibinfo  {journal} {Nature}\ }\textbf {\bibinfo {volume} {592}}
  (\bibinfo {year} {2021}),\ 10.1038/s41586-021-03319-3}\BibitemShut {NoStop}%
\bibitem [{\citenamefont {Sharpe}\ \emph {et~al.}(2019)\citenamefont {Sharpe},
  \citenamefont {Fox}, \citenamefont {Barnard}, \citenamefont {Finney},
  \citenamefont {Watanabe}, \citenamefont {Taniguchi}, \citenamefont
  {Kastner},\ and\ \citenamefont {Goldhaber-Gordon}}]{Sharpe2019}%
  \BibitemOpen
  \bibfield  {author} {\bibinfo {author} {\bibfnamefont {Aaron~L.}\
  \bibnamefont {Sharpe}}, \bibinfo {author} {\bibfnamefont {Eli~J.}\
  \bibnamefont {Fox}}, \bibinfo {author} {\bibfnamefont {Arthur~W.}\
  \bibnamefont {Barnard}}, \bibinfo {author} {\bibfnamefont {Joe}\ \bibnamefont
  {Finney}}, \bibinfo {author} {\bibfnamefont {Kenji}\ \bibnamefont
  {Watanabe}}, \bibinfo {author} {\bibfnamefont {Takashi}\ \bibnamefont
  {Taniguchi}}, \bibinfo {author} {\bibfnamefont {M.~A.}\ \bibnamefont
  {Kastner}}, \ and\ \bibinfo {author} {\bibfnamefont {David}\ \bibnamefont
  {Goldhaber-Gordon}},\ }\bibfield  {title} {\enquote {\bibinfo {title}
  {{Emergent ferromagnetism near three-quarters filling in twisted bilayer
  graphene}},}\ }\href@noop {} {\bibfield  {journal} {\bibinfo  {journal}
  {Science}\ }\textbf {\bibinfo {volume} {608}},\ \bibinfo {pages} {605--608}
  (\bibinfo {year} {2019})}\BibitemShut {NoStop}%
\bibitem [{\citenamefont {Park}\ \emph {et~al.}(2021)\citenamefont {Park},
  \citenamefont {Cao}, \citenamefont {Watanabe}, \citenamefont {Taniguchi},\
  and\ \citenamefont {Jarillo-Herrero}}]{Park2021}%
  \BibitemOpen
  \bibfield  {author} {\bibinfo {author} {\bibfnamefont {Jeong~Min}\
  \bibnamefont {Park}}, \bibinfo {author} {\bibfnamefont {Yuan}\ \bibnamefont
  {Cao}}, \bibinfo {author} {\bibfnamefont {Kenji}\ \bibnamefont {Watanabe}},
  \bibinfo {author} {\bibfnamefont {Takashi}\ \bibnamefont {Taniguchi}}, \ and\
  \bibinfo {author} {\bibfnamefont {Pablo}\ \bibnamefont {Jarillo-Herrero}},\
  }\bibfield  {title} {\enquote {\bibinfo {title} {{Flavour Hund's coupling,
  Chern gaps and charge diffusivity in moir{\'{e}} graphene}},}\ }\href
  {\doibase 10.1038/s41586-021-03366-w} {\bibfield  {journal} {\bibinfo
  {journal} {Nature}\ }\textbf {\bibinfo {volume} {592}},\ \bibinfo {pages}
  {43} (\bibinfo {year} {2021})}\BibitemShut {NoStop}%
\bibitem [{\citenamefont {Jiang}\ \emph {et~al.}(2019)\citenamefont {Jiang},
  \citenamefont {Lai}, \citenamefont {Watanabe}, \citenamefont {Taniguchi},
  \citenamefont {Haule}, \citenamefont {Mao},\ and\ \citenamefont
  {Andrei}}]{Jiang2019}%
  \BibitemOpen
  \bibfield  {author} {\bibinfo {author} {\bibfnamefont {Yuhang}\ \bibnamefont
  {Jiang}}, \bibinfo {author} {\bibfnamefont {Xinyuan}\ \bibnamefont {Lai}},
  \bibinfo {author} {\bibfnamefont {Kenji}\ \bibnamefont {Watanabe}}, \bibinfo
  {author} {\bibfnamefont {Takashi}\ \bibnamefont {Taniguchi}}, \bibinfo
  {author} {\bibfnamefont {Kristjan}\ \bibnamefont {Haule}}, \bibinfo {author}
  {\bibfnamefont {Jinhai}\ \bibnamefont {Mao}}, \ and\ \bibinfo {author}
  {\bibfnamefont {Eva~Y.}\ \bibnamefont {Andrei}},\ }\bibfield  {title}
  {\enquote {\bibinfo {title} {{Charge order and broken rotational symmetry in
  magic-angle twisted bilayer graphene}},}\ }\href {\doibase
  10.1038/s41586-019-1460-4} {\bibfield  {journal} {\bibinfo  {journal}
  {Nature}\ }\textbf {\bibinfo {volume} {573}},\ \bibinfo {pages} {91--95}
  (\bibinfo {year} {2019})}\BibitemShut {NoStop}%
\bibitem [{\citenamefont {Kerelsky}\ \emph {et~al.}(2019)\citenamefont
  {Kerelsky}, \citenamefont {McGilly}, \citenamefont {Kennes}, \citenamefont
  {Xian}, \citenamefont {Yankowitz}, \citenamefont {Chen}, \citenamefont
  {Watanabe}, \citenamefont {Taniguchi}, \citenamefont {Hone}, \citenamefont
  {Dean}, \citenamefont {Rubio},\ and\ \citenamefont
  {Pasupathy}}]{Kerelsky2019}%
  \BibitemOpen
  \bibfield  {author} {\bibinfo {author} {\bibfnamefont {Alexander}\
  \bibnamefont {Kerelsky}}, \bibinfo {author} {\bibfnamefont {Leo~J.}\
  \bibnamefont {McGilly}}, \bibinfo {author} {\bibfnamefont {Dante~M.}\
  \bibnamefont {Kennes}}, \bibinfo {author} {\bibfnamefont {Lede}\ \bibnamefont
  {Xian}}, \bibinfo {author} {\bibfnamefont {Matthew}\ \bibnamefont
  {Yankowitz}}, \bibinfo {author} {\bibfnamefont {Shaowen}\ \bibnamefont
  {Chen}}, \bibinfo {author} {\bibfnamefont {K.}~\bibnamefont {Watanabe}},
  \bibinfo {author} {\bibfnamefont {T.}~\bibnamefont {Taniguchi}}, \bibinfo
  {author} {\bibfnamefont {James}\ \bibnamefont {Hone}}, \bibinfo {author}
  {\bibfnamefont {Cory}\ \bibnamefont {Dean}}, \bibinfo {author} {\bibfnamefont
  {Angel}\ \bibnamefont {Rubio}}, \ and\ \bibinfo {author} {\bibfnamefont
  {Abhay~N.}\ \bibnamefont {Pasupathy}},\ }\bibfield  {title} {\enquote
  {\bibinfo {title} {{Maximized electron interactions at the magic angle in
  twisted bilayer graphene}},}\ }\href {\doibase 10.1038/s41586-019-1431-9}
  {\bibfield  {journal} {\bibinfo  {journal} {Nature}\ }\textbf {\bibinfo
  {volume} {572}},\ \bibinfo {pages} {95--100} (\bibinfo {year}
  {2019})}\BibitemShut {NoStop}%
\bibitem [{\citenamefont {Lu}\ \emph {et~al.}(2019)\citenamefont {Lu},
  \citenamefont {Stepanov}, \citenamefont {Yang}, \citenamefont {Xie},
  \citenamefont {Aamir}, \citenamefont {Das}, \citenamefont {Urgell},
  \citenamefont {Watanabe}, \citenamefont {Taniguchi}, \citenamefont {Zhang},
  \citenamefont {Bachtold}, \citenamefont {MacDonald},\ and\ \citenamefont
  {Efetov}}]{Lu2019}%
  \BibitemOpen
  \bibfield  {author} {\bibinfo {author} {\bibfnamefont {Xiaobo}\ \bibnamefont
  {Lu}}, \bibinfo {author} {\bibfnamefont {Petr}\ \bibnamefont {Stepanov}},
  \bibinfo {author} {\bibfnamefont {Wei}\ \bibnamefont {Yang}}, \bibinfo
  {author} {\bibfnamefont {Ming}\ \bibnamefont {Xie}}, \bibinfo {author}
  {\bibfnamefont {Mohammed~Ali}\ \bibnamefont {Aamir}}, \bibinfo {author}
  {\bibfnamefont {Ipsita}\ \bibnamefont {Das}}, \bibinfo {author}
  {\bibfnamefont {Carles}\ \bibnamefont {Urgell}}, \bibinfo {author}
  {\bibfnamefont {Kenji}\ \bibnamefont {Watanabe}}, \bibinfo {author}
  {\bibfnamefont {Takashi}\ \bibnamefont {Taniguchi}}, \bibinfo {author}
  {\bibfnamefont {Guangyu}\ \bibnamefont {Zhang}}, \bibinfo {author}
  {\bibfnamefont {Adrian}\ \bibnamefont {Bachtold}}, \bibinfo {author}
  {\bibfnamefont {Allan~H.}\ \bibnamefont {MacDonald}}, \ and\ \bibinfo
  {author} {\bibfnamefont {Dmitri~K.}\ \bibnamefont {Efetov}},\ }\bibfield
  {title} {\enquote {\bibinfo {title} {{Superconductors, orbital magnets and
  correlated states in magic-angle bilayer graphene}},}\ }\href {\doibase
  10.1038/s41586-019-1695-0} {\bibfield  {journal} {\bibinfo  {journal}
  {Nature}\ }\textbf {\bibinfo {volume} {574}},\ \bibinfo {pages} {653--657}
  (\bibinfo {year} {2019})}\BibitemShut {NoStop}%
\bibitem [{\citenamefont {Zondiner}\ \emph {et~al.}(2020)\citenamefont
  {Zondiner}, \citenamefont {Rozen}, \citenamefont {Rodan-Legrain},
  \citenamefont {Cao}, \citenamefont {Queiroz}, \citenamefont {Taniguchi},
  \citenamefont {Watanabe}, \citenamefont {Oreg}, \citenamefont {von Oppen},
  \citenamefont {Stern}, \citenamefont {Berg}, \citenamefont
  {Jarillo-Herrero},\ and\ \citenamefont {Ilani}}]{Zondiner2020}%
  \BibitemOpen
  \bibfield  {author} {\bibinfo {author} {\bibfnamefont {U.}~\bibnamefont
  {Zondiner}}, \bibinfo {author} {\bibfnamefont {A.}~\bibnamefont {Rozen}},
  \bibinfo {author} {\bibfnamefont {D.}~\bibnamefont {Rodan-Legrain}}, \bibinfo
  {author} {\bibfnamefont {Y.}~\bibnamefont {Cao}}, \bibinfo {author}
  {\bibfnamefont {R.}~\bibnamefont {Queiroz}}, \bibinfo {author} {\bibfnamefont
  {T.}~\bibnamefont {Taniguchi}}, \bibinfo {author} {\bibfnamefont
  {K.}~\bibnamefont {Watanabe}}, \bibinfo {author} {\bibfnamefont
  {Y.}~\bibnamefont {Oreg}}, \bibinfo {author} {\bibfnamefont {F.}~\bibnamefont
  {von Oppen}}, \bibinfo {author} {\bibfnamefont {Ady}\ \bibnamefont {Stern}},
  \bibinfo {author} {\bibfnamefont {E.}~\bibnamefont {Berg}}, \bibinfo {author}
  {\bibfnamefont {P.}~\bibnamefont {Jarillo-Herrero}}, \ and\ \bibinfo {author}
  {\bibfnamefont {S.}~\bibnamefont {Ilani}},\ }\bibfield  {title} {\enquote
  {\bibinfo {title} {{Cascade of phase transitions and Dirac revivals in
  magic-angle graphene}},}\ }\href {\doibase 10.1038/s41586-020-2373-y}
  {\bibfield  {journal} {\bibinfo  {journal} {Nature}\ }\textbf {\bibinfo
  {volume} {582}},\ \bibinfo {pages} {203--208} (\bibinfo {year} {2020})},\
  \Eprint {http://arxiv.org/abs/1912.06150} {arXiv:1912.06150} \BibitemShut
  {NoStop}%
\bibitem [{\citenamefont {Serlin}\ \emph {et~al.}(2020)\citenamefont {Serlin},
  \citenamefont {Tschirhart}, \citenamefont {Polshyn}, \citenamefont {Zhang},
  \citenamefont {Zhu}, \citenamefont {Watanabe}, \citenamefont {Taniguchi},
  \citenamefont {Balents},\ and\ \citenamefont {Young}}]{Serlin2020}%
  \BibitemOpen
  \bibfield  {author} {\bibinfo {author} {\bibfnamefont {M.}~\bibnamefont
  {Serlin}}, \bibinfo {author} {\bibfnamefont {C.~L.}\ \bibnamefont
  {Tschirhart}}, \bibinfo {author} {\bibfnamefont {H.}~\bibnamefont {Polshyn}},
  \bibinfo {author} {\bibfnamefont {Y.}~\bibnamefont {Zhang}}, \bibinfo
  {author} {\bibfnamefont {J.}~\bibnamefont {Zhu}}, \bibinfo {author}
  {\bibfnamefont {K.}~\bibnamefont {Watanabe}}, \bibinfo {author}
  {\bibfnamefont {T.}~\bibnamefont {Taniguchi}}, \bibinfo {author}
  {\bibfnamefont {L.}~\bibnamefont {Balents}}, \ and\ \bibinfo {author}
  {\bibfnamefont {A.~F.}\ \bibnamefont {Young}},\ }\bibfield  {title} {\enquote
  {\bibinfo {title} {{Intrinsic quantized anomalous Hall effect in a
  moir{\'{e}} heterostructure}},}\ }\href {\doibase 10.1126/science.aay5533}
  {\bibfield  {journal} {\bibinfo  {journal} {Science}\ }\textbf {\bibinfo
  {volume} {367}},\ \bibinfo {pages} {900--903} (\bibinfo {year} {2020})},\
  \Eprint {http://arxiv.org/abs/1907.00261} {arXiv:1907.00261} \BibitemShut
  {NoStop}%
\bibitem [{\citenamefont {Cao}\ \emph {et~al.}(2021)\citenamefont {Cao},
  \citenamefont {Rodan-Legrain}, \citenamefont {Park}, \citenamefont {Yuan},
  \citenamefont {Watanabe}, \citenamefont {Taniguchi}, \citenamefont
  {Fernandes}, \citenamefont {Fu},\ and\ \citenamefont
  {Jarillo-Herrero}}]{Cao2021}%
  \BibitemOpen
  \bibfield  {author} {\bibinfo {author} {\bibfnamefont {Yuan}\ \bibnamefont
  {Cao}}, \bibinfo {author} {\bibfnamefont {Daniel}\ \bibnamefont
  {Rodan-Legrain}}, \bibinfo {author} {\bibfnamefont {Jeong~Min}\ \bibnamefont
  {Park}}, \bibinfo {author} {\bibfnamefont {Noah~F.Q.}\ \bibnamefont {Yuan}},
  \bibinfo {author} {\bibfnamefont {Kenji}\ \bibnamefont {Watanabe}}, \bibinfo
  {author} {\bibfnamefont {Takashi}\ \bibnamefont {Taniguchi}}, \bibinfo
  {author} {\bibfnamefont {Rafael~M.}\ \bibnamefont {Fernandes}}, \bibinfo
  {author} {\bibfnamefont {Liang}\ \bibnamefont {Fu}}, \ and\ \bibinfo {author}
  {\bibfnamefont {Pablo}\ \bibnamefont {Jarillo-Herrero}},\ }\bibfield  {title}
  {\enquote {\bibinfo {title} {{Nematicity and competing orders in
  superconducting magic-angle graphene}},}\ }\href {\doibase
  10.1126/SCIENCE.ABC2836} {\bibfield  {journal} {\bibinfo  {journal}
  {Science}\ }\textbf {\bibinfo {volume} {372}},\ \bibinfo {pages} {264--271}
  (\bibinfo {year} {2021})}\BibitemShut {NoStop}%
\bibitem [{\citenamefont {Tschirhart}\ \emph {et~al.}(2021)\citenamefont
  {Tschirhart}, \citenamefont {Serlin}, \citenamefont {Polshyn}, \citenamefont
  {Shragai}, \citenamefont {Xia}, \citenamefont {Zhu}, \citenamefont {Zhang},
  \citenamefont {Watanabe}, \citenamefont {Taniguchi}, \citenamefont {Huber},\
  and\ \citenamefont {Young}}]{Tschirhart2021}%
  \BibitemOpen
  \bibfield  {author} {\bibinfo {author} {\bibfnamefont {C.~L.}\ \bibnamefont
  {Tschirhart}}, \bibinfo {author} {\bibfnamefont {M.}~\bibnamefont {Serlin}},
  \bibinfo {author} {\bibfnamefont {H.}~\bibnamefont {Polshyn}}, \bibinfo
  {author} {\bibfnamefont {A.}~\bibnamefont {Shragai}}, \bibinfo {author}
  {\bibfnamefont {Z.}~\bibnamefont {Xia}}, \bibinfo {author} {\bibfnamefont
  {J.}~\bibnamefont {Zhu}}, \bibinfo {author} {\bibfnamefont {Y.}~\bibnamefont
  {Zhang}}, \bibinfo {author} {\bibfnamefont {K.}~\bibnamefont {Watanabe}},
  \bibinfo {author} {\bibfnamefont {T.}~\bibnamefont {Taniguchi}}, \bibinfo
  {author} {\bibfnamefont {M.~E.}\ \bibnamefont {Huber}}, \ and\ \bibinfo
  {author} {\bibfnamefont {A.~F.}\ \bibnamefont {Young}},\ }\bibfield  {title}
  {\enquote {\bibinfo {title} {{Imaging orbital ferromagnetism in a moir{\'{e}}
  Chern insulator}},}\ }\href {\doibase 10.1126/SCIENCE.ABD3190/FORMAT/PDF}
  {\bibfield  {journal} {\bibinfo  {journal} {Science}\ }\textbf {\bibinfo
  {volume} {372}},\ \bibinfo {pages} {1323--1327} (\bibinfo {year} {2021})},\
  \Eprint {http://arxiv.org/abs/2006.08053} {arXiv:2006.08053} \BibitemShut
  {NoStop}%
\bibitem [{\citenamefont {Saito}\ \emph {et~al.}(2021)\citenamefont {Saito},
  \citenamefont {Yang}, \citenamefont {Ge}, \citenamefont {Liu}, \citenamefont
  {Taniguchi}, \citenamefont {Watanabe}, \citenamefont {{A Li}}, \citenamefont
  {Berg},\ and\ \citenamefont {Young}}]{Saito2021}%
  \BibitemOpen
  \bibfield  {author} {\bibinfo {author} {\bibfnamefont {Yu}~\bibnamefont
  {Saito}}, \bibinfo {author} {\bibfnamefont {Fangyuan}\ \bibnamefont {Yang}},
  \bibinfo {author} {\bibfnamefont {Jingyuan}\ \bibnamefont {Ge}}, \bibinfo
  {author} {\bibfnamefont {Xiaoxue}\ \bibnamefont {Liu}}, \bibinfo {author}
  {\bibfnamefont {Takashi}\ \bibnamefont {Taniguchi}}, \bibinfo {author}
  {\bibfnamefont {Kenji}\ \bibnamefont {Watanabe}}, \bibinfo {author}
  {\bibfnamefont {J~I}\ \bibnamefont {{A Li}}}, \bibinfo {author}
  {\bibfnamefont {Erez}\ \bibnamefont {Berg}}, \ and\ \bibinfo {author}
  {\bibfnamefont {Andrea~F}\ \bibnamefont {Young}},\ }\bibfield  {title}
  {\enquote {\bibinfo {title} {{Isospin Pomeranchuk effect in twisted bilayer
  graphene}},}\ }\href {\doibase 10.1038/s41586-021-03409-2} {\bibfield
  {journal} {\bibinfo  {journal} {Nature}\ }\textbf {\bibinfo {volume} {592}}
  (\bibinfo {year} {2021}),\ 10.1038/s41586-021-03409-2}\BibitemShut {NoStop}%
\bibitem [{\citenamefont {Breit}(1928)}]{Breit1928}%
  \BibitemOpen
  \bibfield  {author} {\bibinfo {author} {\bibfnamefont {Gregory}\ \bibnamefont
  {Breit}},\ }\bibfield  {title} {\enquote {\bibinfo {title} {{An
  Interpretation of Dirac's Theory of the Electron}},}\ }\href {\doibase
  10.1073/pnas.14.7.553} {\bibfield  {journal} {\bibinfo  {journal}
  {Proceedings of the National Academy of Sciences}\ }\textbf {\bibinfo
  {volume} {14}},\ \bibinfo {pages} {553--559} (\bibinfo {year}
  {1928})}\BibitemShut {NoStop}%
\bibitem [{\citenamefont {Kazmierczak}\ and\ \citenamefont {{Van
  Winkle}}(2021)}]{Kazmierczak}%
  \BibitemOpen
  \bibfield  {author} {\bibinfo {author} {\bibfnamefont {Nathanael~P}\
  \bibnamefont {Kazmierczak}}\ and\ \bibinfo {author} {\bibfnamefont
  {Madeline}\ \bibnamefont {{Van Winkle}}},\ }\bibfield  {title} {\enquote
  {\bibinfo {title} {{Strain fields in twisted bilayer graphene}},}\ }\href
  {\doibase 10.1038/s41563-021-00973-w} {\bibfield  {journal} {\bibinfo
  {journal} {Nature Materials}\ } (\bibinfo {year} {2021}),\
  10.1038/s41563-021-00973-w}\BibitemShut {NoStop}%
\bibitem [{\citenamefont {Bi}\ \emph {et~al.}(2019)\citenamefont {Bi},
  \citenamefont {Yuan},\ and\ \citenamefont {Fu}}]{Bi2019}%
  \BibitemOpen
  \bibfield  {author} {\bibinfo {author} {\bibfnamefont {Zhen}\ \bibnamefont
  {Bi}}, \bibinfo {author} {\bibfnamefont {Noah F~Q}\ \bibnamefont {Yuan}}, \
  and\ \bibinfo {author} {\bibfnamefont {Liang}\ \bibnamefont {Fu}},\
  }\bibfield  {title} {\enquote {\bibinfo {title} {{Designing flat bands by
  strain}},}\ }\href {\doibase 10.1103/PhysRevB.100.035448} {\bibfield
  {journal} {\bibinfo  {journal} {Physical Review B}\ }\textbf {\bibinfo
  {volume} {100}},\ \bibinfo {pages} {35448} (\bibinfo {year}
  {2019})}\BibitemShut {NoStop}%
\bibitem [{\citenamefont {Balents}(2019)}]{balents2019general}%
  \BibitemOpen
  \bibfield  {author} {\bibinfo {author} {\bibfnamefont {Leon}\ \bibnamefont
  {Balents}},\ }\bibfield  {title} {\enquote {\bibinfo {title} {General
  continuum model for twisted bilayer graphene and arbitrary smooth
  deformations},}\ }\href {https://scipost.org/10.21468/SciPostPhys.7.4.048}
  {\bibfield  {journal} {\bibinfo  {journal} {SciPost Phys}\ }\textbf {\bibinfo
  {volume} {7}},\ \bibinfo {pages} {48} (\bibinfo {year} {2019})}\BibitemShut
  {NoStop}%
\bibitem [{\citenamefont {Suzuura}\ and\ \citenamefont
  {Ando}(2002)}]{Suzuura2002}%
  \BibitemOpen
  \bibfield  {author} {\bibinfo {author} {\bibfnamefont {Hidekatsu}\
  \bibnamefont {Suzuura}}\ and\ \bibinfo {author} {\bibfnamefont {Tsuneya}\
  \bibnamefont {Ando}},\ }\bibfield  {title} {\enquote {\bibinfo {title}
  {{Phonons and electron-phonon scattering in carbon nanotubes}},}\ }\href
  {\doibase 10.1103/PhysRevB.65.235412} {\bibfield  {journal} {\bibinfo
  {journal} {Physical Review B - Condensed Matter and Materials Physics}\
  }\textbf {\bibinfo {volume} {65}},\ \bibinfo {pages} {1--15} (\bibinfo {year}
  {2002})}\BibitemShut {NoStop}%
\bibitem [{\citenamefont {{Castro Neto}}\ \emph {et~al.}(2009)\citenamefont
  {{Castro Neto}}, \citenamefont {Guinea}, \citenamefont {Peres}, \citenamefont
  {Novoselov},\ and\ \citenamefont {Geim}}]{CastroNeto2009}%
  \BibitemOpen
  \bibfield  {author} {\bibinfo {author} {\bibfnamefont {A.~H.}\ \bibnamefont
  {{Castro Neto}}}, \bibinfo {author} {\bibfnamefont {F.}~\bibnamefont
  {Guinea}}, \bibinfo {author} {\bibfnamefont {N.~M.R.}\ \bibnamefont {Peres}},
  \bibinfo {author} {\bibfnamefont {K.~S.}\ \bibnamefont {Novoselov}}, \ and\
  \bibinfo {author} {\bibfnamefont {A.~K.}\ \bibnamefont {Geim}},\ }\bibfield
  {title} {\enquote {\bibinfo {title} {{The electronic properties of
  graphene}},}\ }\href {\doibase 10.1103/RevModPhys.81.109} {\bibfield
  {journal} {\bibinfo  {journal} {Reviews of Modern Physics}\ }\textbf
  {\bibinfo {volume} {81}},\ \bibinfo {pages} {109--162} (\bibinfo {year}
  {2009})},\ \Eprint {http://arxiv.org/abs/0709.1163} {arXiv:0709.1163}
  \BibitemShut {NoStop}%
\bibitem [{\citenamefont {Nam}\ and\ \citenamefont {Koshino}(2017)}]{Nam2017}%
  \BibitemOpen
  \bibfield  {author} {\bibinfo {author} {\bibfnamefont {Nguyen N~T}\
  \bibnamefont {Nam}}\ and\ \bibinfo {author} {\bibfnamefont {Mikito}\
  \bibnamefont {Koshino}},\ }\bibfield  {title} {\enquote {\bibinfo {title}
  {{Lattice relaxation and energy band modulation in twisted bilayer
  graphene}},}\ }\href {\doibase 10.1103/PhysRevB.96.075311} {\bibfield
  {journal} {\bibinfo  {journal} {Physical Review B}\ }\textbf {\bibinfo
  {volume} {96}},\ \bibinfo {pages} {75311} (\bibinfo {year}
  {2017})}\BibitemShut {NoStop}%
\bibitem [{\citenamefont {Koshino}\ \emph {et~al.}(2018)\citenamefont
  {Koshino}, \citenamefont {Yuan}, \citenamefont {Koretsune}, \citenamefont
  {Ochi}, \citenamefont {Kuroki},\ and\ \citenamefont {Fu}}]{Koshino2018}%
  \BibitemOpen
  \bibfield  {author} {\bibinfo {author} {\bibfnamefont {Mikito}\ \bibnamefont
  {Koshino}}, \bibinfo {author} {\bibfnamefont {Noah F~Q}\ \bibnamefont
  {Yuan}}, \bibinfo {author} {\bibfnamefont {Takashi}\ \bibnamefont
  {Koretsune}}, \bibinfo {author} {\bibfnamefont {Masayuki}\ \bibnamefont
  {Ochi}}, \bibinfo {author} {\bibfnamefont {Kazuhiko}\ \bibnamefont {Kuroki}},
  \ and\ \bibinfo {author} {\bibfnamefont {Liang}\ \bibnamefont {Fu}},\
  }\bibfield  {title} {\enquote {\bibinfo {title} {{Maximally Localized Wannier
  Orbitals and the Extended Hubbard Model for Twisted Bilayer Graphene}},}\
  }\href {\doibase 10.1103/PhysRevX.8.031087} {\bibfield  {journal} {\bibinfo
  {journal} {Physical Review X}\ } (\bibinfo {year} {2018}),\
  10.1103/PhysRevX.8.031087}\BibitemShut {NoStop}%
\bibitem [{\citenamefont {Carr}\ \emph {et~al.}(2019)\citenamefont {Carr},
  \citenamefont {Fang}, \citenamefont {Zhu},\ and\ \citenamefont
  {Kaxiras}}]{Carr2019}%
  \BibitemOpen
  \bibfield  {author} {\bibinfo {author} {\bibfnamefont {Stephen}\ \bibnamefont
  {Carr}}, \bibinfo {author} {\bibfnamefont {Shiang}\ \bibnamefont {Fang}},
  \bibinfo {author} {\bibfnamefont {Ziyan}\ \bibnamefont {Zhu}}, \ and\
  \bibinfo {author} {\bibfnamefont {Efthimios}\ \bibnamefont {Kaxiras}},\
  }\bibfield  {title} {\enquote {\bibinfo {title} {Exact continuum model for
  low-energy electronic states of twisted bilayer graphene},}\ }\href {\doibase
  10.1103/PhysRevResearch.1.013001} {\bibfield  {journal} {\bibinfo  {journal}
  {Phys. Rev. Research}\ }\textbf {\bibinfo {volume} {1}},\ \bibinfo {pages}
  {013001} (\bibinfo {year} {2019})}\BibitemShut {NoStop}%
\bibitem [{\citenamefont {Huder}\ \emph {et~al.}(2018)\citenamefont {Huder},
  \citenamefont {Artaud}, \citenamefont {{Le Quang}}, \citenamefont {{De
  Laissardi{\`{e}}re}}, \citenamefont {Jansen}, \citenamefont {Lapertot},
  \citenamefont {Chapelier},\ and\ \citenamefont {Renard}}]{Huder2018}%
  \BibitemOpen
  \bibfield  {author} {\bibinfo {author} {\bibfnamefont {Lo{\"{i}}c}\
  \bibnamefont {Huder}}, \bibinfo {author} {\bibfnamefont {Alexandre}\
  \bibnamefont {Artaud}}, \bibinfo {author} {\bibfnamefont {Toai}\ \bibnamefont
  {{Le Quang}}}, \bibinfo {author} {\bibfnamefont {Guy~Trambly}\ \bibnamefont
  {{De Laissardi{\`{e}}re}}}, \bibinfo {author} {\bibfnamefont {Aloysius~G.M.}\
  \bibnamefont {Jansen}}, \bibinfo {author} {\bibfnamefont {G{\'{e}}rard}\
  \bibnamefont {Lapertot}}, \bibinfo {author} {\bibfnamefont {Claude}\
  \bibnamefont {Chapelier}}, \ and\ \bibinfo {author} {\bibfnamefont
  {Vincent~T.}\ \bibnamefont {Renard}},\ }\bibfield  {title} {\enquote
  {\bibinfo {title} {{Electronic Spectrum of Twisted Graphene Layers under
  Heterostrain}},}\ }\href {\doibase 10.1103/PhysRevLett.120.156405} {\bibfield
   {journal} {\bibinfo  {journal} {Physical Review Letters}\ }\textbf {\bibinfo
  {volume} {120}},\ \bibinfo {pages} {1--5} (\bibinfo {year} {2018})},\ \Eprint
  {http://arxiv.org/abs/1803.03505} {arXiv:1803.03505} \BibitemShut {NoStop}%
\bibitem [{\citenamefont {Qiao}\ \emph {et~al.}(2018)\citenamefont {Qiao},
  \citenamefont {Yin},\ and\ \citenamefont {He}}]{Qiao2018}%
  \BibitemOpen
  \bibfield  {author} {\bibinfo {author} {\bibfnamefont {Jia~Bin}\ \bibnamefont
  {Qiao}}, \bibinfo {author} {\bibfnamefont {Long~Jing}\ \bibnamefont {Yin}}, \
  and\ \bibinfo {author} {\bibfnamefont {Lin}\ \bibnamefont {He}},\ }\bibfield
  {title} {\enquote {\bibinfo {title} {{Twisted graphene bilayer around the
  first magic angle engineered by heterostrain}},}\ }\href {\doibase
  10.1103/PhysRevB.98.235402} {\bibfield  {journal} {\bibinfo  {journal}
  {Physical Review B}\ }\textbf {\bibinfo {volume} {98}},\ \bibinfo {pages}
  {1--8} (\bibinfo {year} {2018})}\BibitemShut {NoStop}%
\bibitem [{\citenamefont {Liu}\ \emph {et~al.}(2021)\citenamefont {Liu},
  \citenamefont {Khalaf}, \citenamefont {Lee},\ and\ \citenamefont
  {Vishwanath}}]{Liu2021nematic}%
  \BibitemOpen
  \bibfield  {author} {\bibinfo {author} {\bibfnamefont {Shang}\ \bibnamefont
  {Liu}}, \bibinfo {author} {\bibfnamefont {Eslam}\ \bibnamefont {Khalaf}},
  \bibinfo {author} {\bibfnamefont {Jong~Yeon}\ \bibnamefont {Lee}}, \ and\
  \bibinfo {author} {\bibfnamefont {Ashvin}\ \bibnamefont {Vishwanath}},\
  }\bibfield  {title} {\enquote {\bibinfo {title} {{Nematic topological
  semimetal and insulator in magic-angle bilayer graphene at charge
  neutrality}},}\ }\href {\doibase 10.1103/PhysRevResearch.3.013033} {\bibfield
   {journal} {\bibinfo  {journal} {Physical Review Research}\ }\textbf
  {\bibinfo {volume} {3}},\ \bibinfo {pages} {13033} (\bibinfo {year}
  {2021})}\BibitemShut {NoStop}%
\bibitem [{\citenamefont {Parker}\ \emph {et~al.}(2021)\citenamefont {Parker},
  \citenamefont {Soejima}, \citenamefont {Hauschild}, \citenamefont {Zaletel},\
  and\ \citenamefont {Bultinck}}]{Parker2021}%
  \BibitemOpen
  \bibfield  {author} {\bibinfo {author} {\bibfnamefont {Daniel~E}\
  \bibnamefont {Parker}}, \bibinfo {author} {\bibfnamefont {Tomohiro}\
  \bibnamefont {Soejima}}, \bibinfo {author} {\bibfnamefont {Johannes}\
  \bibnamefont {Hauschild}}, \bibinfo {author} {\bibfnamefont {Michael~P}\
  \bibnamefont {Zaletel}}, \ and\ \bibinfo {author} {\bibfnamefont {Nick}\
  \bibnamefont {Bultinck}},\ }\bibfield  {title} {\enquote {\bibinfo {title}
  {{Strain-Induced Quantum Phase Transitions in Magic-Angle Graphene}},}\
  }\href {\doibase 10.1103/PhysRevLett.127.027601} {\bibfield  {journal}
  {\bibinfo  {journal} {Physical Review Letters}\ }\textbf {\bibinfo {volume}
  {127}} (\bibinfo {year} {2021}),\ 10.1103/PhysRevLett.127.027601}\BibitemShut
  {NoStop}%
\bibitem [{\citenamefont {Bernevig}\ \emph {et~al.}(2021)\citenamefont
  {Bernevig}, \citenamefont {Song}, \citenamefont {Regnault},\ and\
  \citenamefont {Lian}}]{Bernevig2021}%
  \BibitemOpen
  \bibfield  {author} {\bibinfo {author} {\bibfnamefont {B~Andrei}\
  \bibnamefont {Bernevig}}, \bibinfo {author} {\bibfnamefont {Zhi-Da}\
  \bibnamefont {Song}}, \bibinfo {author} {\bibfnamefont {Nicolas}\
  \bibnamefont {Regnault}}, \ and\ \bibinfo {author} {\bibfnamefont {Biao}\
  \bibnamefont {Lian}},\ }\bibfield  {title} {\enquote {\bibinfo {title}
  {{Twisted bilayer graphene. I. Matrix elements, approximations, perturbation
  theory, and a $k\cdot p$ two-band model}},}\ }\href {\doibase
  10.1103/PhysRevB.103.205411} {\bibfield  {journal} {\bibinfo  {journal}
  {Physical Review B}\ }\textbf {\bibinfo {volume} {103}},\ \bibinfo {pages}
  {205411} (\bibinfo {year} {2021})}\BibitemShut {NoStop}%
\bibitem [{\citenamefont {Kwan}\ \emph {et~al.}(2020)\citenamefont {Kwan},
  \citenamefont {Parameswaran},\ and\ \citenamefont {Sondhi}}]{Kwan2020}%
  \BibitemOpen
  \bibfield  {author} {\bibinfo {author} {\bibfnamefont {Yves~H.}\ \bibnamefont
  {Kwan}}, \bibinfo {author} {\bibfnamefont {S.~A.}\ \bibnamefont
  {Parameswaran}}, \ and\ \bibinfo {author} {\bibfnamefont {S.~L.}\
  \bibnamefont {Sondhi}},\ }\bibfield  {title} {\enquote {\bibinfo {title}
  {Twisted bilayer graphene in a parallel magnetic field},}\ }\href {\doibase
  10.1103/PhysRevB.101.205116} {\bibfield  {journal} {\bibinfo  {journal}
  {Phys. Rev. B}\ }\textbf {\bibinfo {volume} {101}},\ \bibinfo {pages}
  {205116} (\bibinfo {year} {2020})}\BibitemShut {NoStop}%
\bibitem [{\citenamefont {Xie}\ and\ \citenamefont
  {Macdonald}(2020)}]{Xie2020}%
  \BibitemOpen
  \bibfield  {author} {\bibinfo {author} {\bibfnamefont {Ming}\ \bibnamefont
  {Xie}}\ and\ \bibinfo {author} {\bibfnamefont {A.~H.}\ \bibnamefont
  {Macdonald}},\ }\bibfield  {title} {\enquote {\bibinfo {title} {{Nature of
  the Correlated Insulator States in Twisted Bilayer Graphene}},}\ }\href
  {\doibase 10.1103/PhysRevLett.124.097601} {\bibfield  {journal} {\bibinfo
  {journal} {Physical Review Letters}\ }\textbf {\bibinfo {volume} {124}},\
  \bibinfo {pages} {97601} (\bibinfo {year} {2020})}\BibitemShut {NoStop}%
\bibitem [{\citenamefont {Xie}\ and\ \citenamefont
  {Macdonald}(2021)}]{Xie2021}%
  \BibitemOpen
  \bibfield  {author} {\bibinfo {author} {\bibfnamefont {Ming}\ \bibnamefont
  {Xie}}\ and\ \bibinfo {author} {\bibfnamefont {A~H}\ \bibnamefont
  {Macdonald}},\ }\bibfield  {title} {\enquote {\bibinfo {title} {{Weak-Field
  Hall Resistivity and Spin-Valley Flavor Symmetry Breaking in Magic-Angle
  Twisted Bilayer Graphene}},}\ }\href {\doibase
  10.1103/PhysRevLett.127.196401} {\bibfield  {journal} {\bibinfo  {journal}
  {Physical Review Letters}\ }\textbf {\bibinfo {volume} {127}} (\bibinfo
  {year} {2021}),\ 10.1103/PhysRevLett.127.196401}\BibitemShut {NoStop}%
\bibitem [{\citenamefont {Zhang}\ \emph {et~al.}(2020)\citenamefont {Zhang},
  \citenamefont {Hou}, \citenamefont {Zhao}, \citenamefont {Guo}, \citenamefont
  {Liu}, \citenamefont {Li}, \citenamefont {Ren}, \citenamefont {Sun},\ and\
  \citenamefont {He}}]{Zhang2020a}%
  \BibitemOpen
  \bibfield  {author} {\bibinfo {author} {\bibfnamefont {Yu}~\bibnamefont
  {Zhang}}, \bibinfo {author} {\bibfnamefont {Zhe}\ \bibnamefont {Hou}},
  \bibinfo {author} {\bibfnamefont {Ya-Xin~Xin}\ \bibnamefont {Zhao}}, \bibinfo
  {author} {\bibfnamefont {Zi-Han~Han}\ \bibnamefont {Guo}}, \bibinfo {author}
  {\bibfnamefont {Yi-Wen~Wen}\ \bibnamefont {Liu}}, \bibinfo {author}
  {\bibfnamefont {Si-Yu~Yu}\ \bibnamefont {Li}}, \bibinfo {author}
  {\bibfnamefont {Ya-Ning~Ning}\ \bibnamefont {Ren}}, \bibinfo {author}
  {\bibfnamefont {Qing-Feng~Feng}\ \bibnamefont {Sun}}, \ and\ \bibinfo
  {author} {\bibfnamefont {Lin}\ \bibnamefont {He}},\ }\bibfield  {title}
  {\enquote {\bibinfo {title} {{Correlation-induced valley splitting and
  orbital magnetism in a strain-induced zero-energy flatband in twisted bilayer
  graphene near the magic angle}},}\ }\href {\doibase
  10.1103/PhysRevB.102.081403} {\bibfield  {journal} {\bibinfo  {journal}
  {Physical Review B}\ }\textbf {\bibinfo {volume} {102}} (\bibinfo {year}
  {2020}),\ 10.1103/PhysRevB.102.081403}\BibitemShut {NoStop}%
\bibitem [{\citenamefont {He}\ \emph {et~al.}(2020)\citenamefont {He},
  \citenamefont {Goldhaber-Gordon},\ and\ \citenamefont {Law}}]{He2020}%
  \BibitemOpen
  \bibfield  {author} {\bibinfo {author} {\bibfnamefont {Wen~Yu}\ \bibnamefont
  {He}}, \bibinfo {author} {\bibfnamefont {David}\ \bibnamefont
  {Goldhaber-Gordon}}, \ and\ \bibinfo {author} {\bibfnamefont {K.~T.}\
  \bibnamefont {Law}},\ }\bibfield  {title} {\enquote {\bibinfo {title} {{Giant
  orbital magnetoelectric effect and current-induced magnetization switching in
  twisted bilayer graphene}},}\ }\href {\doibase 10.1038/s41467-020-15473-9}
  {\bibfield  {journal} {\bibinfo  {journal} {Nature Communications}\ }\textbf
  {\bibinfo {volume} {11}},\ \bibinfo {pages} {1--8} (\bibinfo {year}
  {2020})},\ \Eprint {http://arxiv.org/abs/1908.11718} {arXiv:1908.11718}
  \BibitemShut {NoStop}%
\bibitem [{\citenamefont {Bultinck}\ \emph {et~al.}(2020)\citenamefont
  {Bultinck}, \citenamefont {Chatterjee},\ and\ \citenamefont
  {Zaletel}}]{Bultinck2020}%
  \BibitemOpen
  \bibfield  {author} {\bibinfo {author} {\bibfnamefont {Nick}\ \bibnamefont
  {Bultinck}}, \bibinfo {author} {\bibfnamefont {Shubhayu}\ \bibnamefont
  {Chatterjee}}, \ and\ \bibinfo {author} {\bibfnamefont {Michael~P}\
  \bibnamefont {Zaletel}},\ }\bibfield  {title} {\enquote {\bibinfo {title}
  {{Mechanism for Anomalous Hall Ferromagnetism in Twisted Bilayer
  Graphene}},}\ }\href {\doibase 10.1103/PhysRevLett.124.166601} {\bibfield
  {journal} {\bibinfo  {journal} {Physical Review Letters}\ }\textbf {\bibinfo
  {volume} {124}} (\bibinfo {year} {2020}),\
  10.1103/PhysRevLett.124.166601}\BibitemShut {NoStop}%
\end{thebibliography}%
\appendix

\section{Tripod model with strain}\label{sec:app_tripod}
In this section we derive 
Eqs.~(\ref{eq:tripod_to_first_order}-\ref{eq:tripod_shift_by_strain}) within the ``tripod model''. Our starting point is the following Hamiltonian:
\begin{align}
\begin{split}
    H(\vec{k}) &= h^{(1)}\left(\vec{k}-\frac{e}{\hbar}\vec{A}_{\text{em}}\right)\\
    &-\sum_{j=0,1,2}T_{j}
    \left[
    h^{(2)}\left(\vec{k}+\frac{e}{\hbar}\vec{A}_{\text{em}}+\vec{g}_{j}\right)-E(\vec{k})
    \right]^{-1}T_{j}^{\dagger},
\end{split}
\end{align}
which is valid at small $\vec{p}=\vec{k}-\delta\vec{K}^{(1)}$ and small parallel magnetic field. To simplify our derivation, we consider a shifted momentum $\vec{p}'=\vec{p}+\frac{e}{\hbar}\vec{A}_{\text{em}}$ such that all the magnetic field terms appear on the first layer:
\begin{subequations}
\begin{align}
    &h^{(1)}\left(\vec{k}-\frac{e}{\hbar}\vec{A}_{\text{em}}\right)
    =\hbar v_{F}\left(\vec{p}'-2\frac{e}{\hbar}\vec{A}_{\text{em}}\right)
    \cdot\bm{\sigma},\\
    &h^{(2)}\left(\vec{k}+\frac{e}{\hbar}\vec{A}_{\text{em}}+\vec{g}_{j}\right)
    =\hbar v_{F}(\vec{p}'+\vec{g}_{j}+\delta\vec{K})
    \cdot\bm{\sigma},
\end{align}
\end{subequations}
where $\delta\vec{K}=\delta\vec{K}^{(1)}-\delta\vec{K}^{(2)}$, and we neglected the rotation of the Pauli matrices as we focus on leading order effects.
The reference scale for momentum and energy is set by introducing the following dimensionless parameters:
\begin{equation}
    \tilde{\vec{p}}=\frac{\vec{p}'}{K\theta},\quad\tilde{E}=\frac{E}{\hbar v_{F}K\theta},\quad
    \vec{d}_{j}=\frac{1}{K\theta}\left(\vec{g}_{j}+\delta\vec{K}\right).
\end{equation}
The corresponding dimensionless tunneling is given:
\begin{equation}\label{eq:app_dimensionless_tunneling}
     \tilde{T}_{j}=u+w\bm{\zeta}_{j}\cdot\bm{\sigma},\quad
     \bm{\zeta}_{j}=\left(\text{Re}\left[\zeta^{j}\right],-\text{Im}\left[\zeta^{j}\right]\right),\quad\zeta=e^{i\frac{2\pi}{3}}.
\end{equation}
In terms of these definitions, our tripod model can be rewritten as:
\begin{align}\label{eq:app_dimensionless_tripod_model}
\begin{split}
    \tilde{H}(\tilde{\vec{p}})&=\left(\tilde{\vec{p}}-2\frac{e}{\hbar K\theta}\vec{A}_{\text{em}}\right)\cdot\bm{\sigma}
    \\
    &
    -\sum_{j=0,1,2}\tilde{T}_{j}\left[
    \left(\tilde{\vec{p}}+\vec{d}_{j}\right)\cdot\bm{\sigma}+\tilde{E}(\tilde{\vec{p}})
    \right]^{-1}\tilde{T}_{j}.
\end{split}
\end{align}

We are interested in expanding Eq.~\eqref{eq:app_dimensionless_tripod_model} to first order in $\tilde{\vec{p}}$, $\tilde{E}$ and in components of $\mathcal{S}/\theta$. Specifically, the task at hand is expanding the inverse of the second layer's Hamiltonian. We start by explicitly writing the form of $\vec{d}_{j}$:
\begin{align}\label{eq:app_d_j_defs}
\begin{split}
    \vec{d}_{j} &= \frac{1}{K\theta}\left[\mathcal{U}^{T}\left(\vec{G}_{j}-\vec{K}\right)-\left(\vec{A}_{s}^{(1)}-\vec{A}_{s}^{(2)}\right)\right]
    \\
    &=\frac{1}{\theta}\mathcal{R}(\theta)\bm{\zeta}_{j}-\frac{1}{\theta}\left[\mathcal{S}\bm{\zeta}_{j}+\frac{1}{K}\left(\vec{A}_{s}^{(1)}-\vec{A}_{s}^{(2)}\right)\right]
    \\
    &=\vec{d}_{j}^{(0)}+\vec{d}_{j}^{(1)},
\end{split}
\end{align}
where we used $\vec{K}-\vec{G}_{j} = K\bm{\zeta}_{j}$, and separated the $\mathcal{O}(1)$ contribution, $\vec{d}_{j}^{(0)} = \theta^{-1}\mathcal{R}(\theta)\bm{\zeta}_{j}$, from the $\mathcal{O}(\mathcal{S}/\theta)$ corrections, $\vec{d}_{j}^{(1)} = \vec{d}_{j}-\vec{d}_{j}^{(0)}$. It will be useful later to introduce a dimensionless \textit{artificial} gauge field $\tilde{\vec{A}}_{s}$ such that $\vec{d}_{j}^{(1)}=-\theta^{-1}\mathcal{S}\zeta_{j}-\tilde{\vec{A}}_{s}$.
Next, we invert the operator between the tunneling matrices, and keep only first order correction in strain and momentum. Using $\left|\vec{d}_{j}^{(0)}\right|^2=1$, we find:
\begin{align}\label{eq:app_inverted_2nd_layer}
\begin{split}
    &\left[
    \left(\tilde{\vec{p}}+\vec{d}_{j}\right)\cdot\bm{\sigma}+\tilde{E}
    \right]^{-1}
    \\
    &\approx 
    \left(\tilde{\vec{p}}+\vec{d}_{j}^{(1)}+\left[1-2\vec{d}_{j}^{(0)}\cdot\left(\tilde{\vec{p}}+\vec{d}_{j}^{(1)}\right)\right]\vec{d}_{j}^{(0)}\right)\cdot\bm{\sigma}-\tilde{E}
    \\
    &=\vec{f}_{j}\cdot\bm{\sigma}-\tilde{E},
\end{split}
\end{align}
where the last line defines $\vec{f}_{j}$. 

Using the form of the dimensionless tunneling in Eq.~\eqref{eq:app_dimensionless_tunneling}, we have the following identity:
\begin{align}\label{eq:app_tunneling_identity}
\begin{split}
    \tilde{T}_{j}(\vec{f}_{j}\cdot\bm{\sigma}-\tilde{E})\tilde{T}_{j}&=
    \left[(u^2-w^2)\vec{f}_{j}+2w^2(\vec{f}_{j}\cdot\bm{\zeta}_{j})\bm{\zeta}_{j}\right]\cdot\bm{\sigma}
    \\
    &-2uw\tilde{E}\bm{\zeta}_{j}\cdot\bm{\sigma}+2uw\vec{f}_{j}\cdot\bm{\zeta}_{j}-(u^2+w^2)\tilde{E}.
\end{split}
\end{align}
Noting that the following cancel - $\sum_{j}\bm{\zeta}_{j}=0$, $\sum_{j}\vec{d}_{j}^{(0)}=0$ and $\vec{d}_{j}^{(0)}\cdot\bm{\zeta}_{j}=0$, we insert Eq.~\eqref{eq:app_inverted_2nd_layer} into Eq.~\eqref{eq:app_tunneling_identity} and obtain:
\begin{align}\label{eq:app_tunneling_after_inversion}
\begin{split}
    &\sum_{j}\tilde{T}_{j}(\vec{f}_{j}\cdot\bm{\sigma}-\tilde{E})\tilde{T}_{j}
    \\
    &=
    (u^2-w^2)\sum_{j}\left(\tilde{\vec{p}}+\vec{d}_{j}^{(1)}-2\left[\vec{d}_{j}^{(0)}\cdot\left(\tilde{\vec{p}}+\vec{d}_{j}^{(1)}\right)\right]\vec{d}_{j}^{(0)}\right)\cdot\bm{\sigma} \\
    &+2w^2\sum_{j}((\tilde{\vec{p}}+\vec{d}_{j}^{(1)})\cdot\bm{\zeta}_{j})\bm{\zeta}_{j}\cdot\bm{\sigma}
    \\
    &-3(u^2+w^2)\tilde{E}+2uw\sum_{j}\vec{d}_{j}^{(1)}\cdot\bm{\zeta}_{j}.
\end{split}
\end{align}
Next, we perform the summation over $j$. For brevity, we divide it into two steps. First, we reinstate the explicit form of $\vec{d}_{j}^{(1)}$ given by Eq.~\eqref{eq:app_d_j_defs}. Making use of $\sum_{j}(\tilde{\vec{p}}\cdot\bm{\zeta}_{j})\cdot\bm{\zeta}_{j}=\sum_{j}(\tilde{\vec{p}}\cdot\vec{d}_{j}^{(0)})\cdot\vec{d}_{j}^{(0)}=\frac{3}{2}\tilde{\vec{p}}$, we rewrite Eq.~\eqref{eq:app_tunneling_after_inversion} as:
\begin{align}
\begin{split}
    &\sum_{j}\tilde{T}_{j}(\vec{f}_{j}\cdot\bm{\sigma}-\tilde{E})\tilde{T}_{j}
    \\
    &=
    2(u^2-w^2)\theta^{-1}\sum_{j}\left[\vec{d}_{j}^{(0)}\cdot\left(\mathcal{S}\bm{\zeta}_{j}\right)\right]\vec{d}_{j}^{(0)}\cdot\bm{\sigma} \\
    &+3w^2(\tilde{\vec{p}}-\tilde{\vec{A}}_{s})\cdot\bm{\sigma}-2w^2\theta^{-1}\sum_{j}\left[\left(\mathcal{S}\bm{\zeta}_{j}\right)\cdot\bm{\zeta}_{j}\right]\bm{\zeta}_{j}\cdot\bm{\sigma}
    \\
    &-3(u^2+w^2)\tilde{E}-2uw\theta^{-1}\sum_{j}\left(\mathcal{S}\bm{\zeta}_{j}\right)\cdot\bm{\zeta}_{j}.
\end{split}
\end{align}
Second, we perform the summations involving the vectors $\vec{d}_{j}^{(0)}$, $\bm{\zeta}_{j}$ and the tensor $\mathcal{S}$. These require a bit more algebra, and result in the following expression:
\begin{align}
\begin{split}
    &\sum_{j}\tilde{T}_{j}(\vec{f}_{j}\cdot\bm{\sigma}-\tilde{E})\tilde{T}_{j}
    \\
    &=3w^2(\tilde{\vec{p}}-\tilde{\vec{A}}_{s})\cdot\bm{\sigma}-3(u^2+w^2)\tilde{E}-3uw(s'+s'')\theta^{-1}
    \\
    &-\frac{3}{2}u^2(s'-s'')\theta^{-1}\hat{\vec{A}}_{s}\cdot\bm{\sigma},
\end{split}
\end{align}
where the last term contains a unit vector $\hat{\vec{A}}_{s}$ (with the direction of $\tilde{\vec{A}}_{s}$) that originated from manipulations over $\mathcal{S}$. Using $\tilde{\vec{A}}_{s}=\frac{\beta}{2\theta Ka}(s'-s'')\hat{\vec{A}}_{s}$, we can rewrite the above as: 
\begin{align}\label{eq:app_tunneling_after_summation}
\begin{split}
    &\sum_{j}\tilde{T}_{j}(\vec{f}_{j}\cdot\bm{\sigma}-\tilde{E})\tilde{T}_{j}
    \\
    &=3w^2\left[\tilde{\vec{p}}-\left(1+\frac{u^2}{w^2}\frac{Ka}{\beta}\right)\tilde{\vec{A}}_{s}\right]\cdot\bm{\sigma}
    \\
    &-3(u^2+w^2)\tilde{E}-3uw(s'+s'')\theta^{-1}.
\end{split}
\end{align}
The outcome of this calculation is given by inserting Eq.~\eqref{eq:app_tunneling_after_summation} back into the Eq.~\eqref{eq:app_dimensionless_tripod_model} for the effective Hamiltonian:
\begin{align}\label{eq:app_effective_hamiltonian_dimensionless}
\begin{split}
    &H_{\text{eff}}(\tilde{\vec{p}})=
    \\
    &\frac{\left((1-3w^2)\tilde{\vec{p}}-2\frac{e}{\hbar K\theta}\vec{A}_{\text{em}}+3w^2\left(1+\frac{u^2}{w^2}\frac{Ka}{\beta}\right)\tilde{\vec{A}}_{s}\right)\cdot\bm{\sigma}}{1+3(u^2+w^2)}
    \\
    &+\frac{3uw}{1+3(u^2+w^2)}\frac{s'+s''}{\theta}.
\end{split}
\end{align}
Restoring the dimensions of the parameters and rearranging to emphasize the shift of the Dirac point, Eq.~\eqref{eq:app_effective_hamiltonian_dimensionless} is rewritten as:
\begin{align}\label{eq:app_effective_hamiltonian_dimensions}
\begin{split}
    &H_{\text{eff}}(\vec{p})=
    \\
    &\frac{\hbar v_{F}(1-3w^2)\left(\vec{p}-\frac{3w^2}{1-3w^2}\left(1+\frac{u^2}{w^2}\frac{Ka}{\beta}\right)\vec{A}_{s}\right)\cdot\bm{\sigma}}{1+3(u^2+w^2)}
    \\
    &+\frac{2}{1+3(u^2+w^2)}ev_{F}\vec{A}_{\text{em}}\cdot\bm{\sigma}
    \\
    &+\hbar v_{F}K\frac{3uw}{1+3(u^2+w^2)}(s'+s'').
\end{split}
\end{align}
One can clearly see from the result in  Eq.~\eqref{eq:app_effective_hamiltonian_dimensions} the sensitivity of the crossing point on the magic angle condition $w^2=1/3$ in the denominator, and that the response to the magnetic field is modified due to tunneling by a prefactor that does not vanish at the magic angle.
\section{Symmetry constraints on magnetization for higher orders in strain}\label{sec:app_symmetry}

\subsection{Effect of $C_3$ symmetry on the energy}

For the purpose of deriving the constraints imposed on the energy by the $C_3$ symmetry we can omit the valley fillings as they are conserved. Focusing on the strain and magnetic field, we rewrite \eqref{eq:energy_symmetry_c3} here as:
\begin{equation}\label{eq:app_energy_symmetry_c3}
    E\left(\mathcal{S},\vec{B}_{\parallel}\right)
    =E\left(C_3\mathcal{S}C^T_3,C_3\vec{B}_{\parallel}\right).
\end{equation}
For the zero-field magnetization, we are interested only in terms of Eq.~\eqref{eq:app_energy_symmetry_c3} that are linear in $\vec{B}_{\parallel}$. 

The constraint in Eq.~\eqref{eq:app_energy_symmetry_c3} allows only for combinations of strain tensor components $s_{\alpha\beta}$ and magnetic field vector components $B_{\gamma}$ that remain invariant under $C_{3}$. The transformation laws for these components are given by:
\begin{align}\label{eq:c3_transform_laws}
\begin{split}
    s_{\alpha\beta} &\rightarrow \left[C_{3}\right]_{\alpha\alpha'}\left[C_{3}\right]_{\beta\beta'}s_{\alpha'\beta'} \\
    B_{\gamma} &\rightarrow \left[C_{3}\right]_{\gamma\gamma'}B_{\gamma'}
\end{split}
\end{align}
where the $C_{3}$ rotation tensor components are $\left[C_{3}\right]_{\alpha\alpha'} = \left[e^{i\frac{2\pi}{3}\eta_{y}}\right]_{\alpha\alpha'}$. 

The $C_{3}$ rotation tensor has the following eigenvectors and corresponding eigenvalues:
\begin{align}\label{eq:app_rotation_egvcs}
\begin{split}
    \left[C_{3}\right]_{\alpha\alpha'}(\delta_{\alpha',x}+i\delta_{\alpha',y}) &= e^{i2\pi/3}(\delta_{\alpha,x}+i\delta_{\alpha,y}),
    \\
    \left[C_{3}\right]_{\alpha\alpha'}(\delta_{\alpha',x}-i\delta_{\alpha',y}) &= e^{-i2\pi/3}(\delta_{\alpha,x}-i\delta_{\alpha,y}).
\end{split}
\end{align}
Eq.~\eqref{eq:app_rotation_egvcs} forces the magnetic field to enter our invariant combination through either $B_{x}+iB_{y}$ or $B_{x}-iB_{y}$. Focusing on the former, we have an eigenvalue of $e^{i2\pi/3}$ corresponding to the magnetic field. For the energy to be invariant under $C_3$, the strain field must present an eigenvalue of $e^{-i2\pi/3}$. This implies that a zeroth order term in strain is prohibited from coupling to a linear term in magnetic field, and thus there is no in-plane magnetization without strain. We will limit the discussion from here on to first and second order in strain, but the generalization is trivial. 

To linear order in strain, the required eigenvalue of $e^{-i2\pi/3}$ can be obtained only by the following combination:
\begin{align}
\begin{split}
    (\delta_{\alpha,x}+i\delta_{\alpha,y})(\delta_{\beta,x}+i\delta_{\beta,y})s_{\alpha\beta} &= s_{xx}-s_{yy}+i2s_{xy} \\
    &=(s'-s'')e^{i2\phi_{S}},
\end{split}
\end{align}
where the second line is written in terms of principal strains $s',s''$ and angle $\phi_{S}$ as defined in Eq.~\eqref{eq:strain_tensor_prinicpal_axis}, and elucidates the rotational nature of this term. A similar argument couples $B_{x}-iB_{y}$ to $(s'-s'')e^{-i2\phi_{S}}$. Rewriting the magnetic field as $\vec{B}=|B|\left(\cos{\phi_{B}},\sin{\phi_{B}}\right)$, and demanding the energy to be real, we find that to linear order in strain and magnetic field:
\begin{equation}
    E(\mathcal{S},\vec{B}_{\parallel}) \approx  \lambda_{1}(s'-s'')|B|\cos{\left(2\phi_{S}+\phi_{B}+\phi_{0}\right)},
\end{equation}
for some $\phi_{0}\in\left[0,\pi\right)$ which will be determined by other symmetries [see below Eq. \eqref{eq:M}]. The prefactor $\lambda_{1}$ is the coupling coefficient at this order. 

The second order in strain must also present the same eigenvalue of $e^{-i2\pi/3}$. In contrast with the linear order, we have two ways of obtaining this eigenvalue:
\begin{align}
\begin{split}
    &(\delta_{\alpha,x}+i\delta_{\alpha,y})(\delta_{\beta,x}+i\delta_{\beta,y})(\delta_{\mu,x}+i\delta_{\mu,y})(\delta_{\nu,x}-i\delta_{\nu,y})s_{\alpha\beta}s_{\mu\nu}  \\
    &=\left((s')^2-(s'')^2\right)e^{i2\phi_{S}},
\end{split}
\end{align}
and
\begin{align}
\begin{split}
    &(\delta_{\alpha,x}-i\delta_{\alpha,y})(\delta_{\beta,x}-i\delta_{\beta,y})(\delta_{\mu,x}-i\delta_{\mu,y})(\delta_{\nu,x}-i\delta_{\nu,y})s_{\alpha\beta}s_{\mu\nu}  \\
    &=\left(s'-s''\right)^2e^{-i4\phi_{S}}.
\end{split}
\end{align}
Following similar arguments as with the linear order in strain, we find the energy to have the form:
\begin{align}\label{eq:app_energy_only_c3}
\begin{split}
    E(\mathcal{S},\vec{B}_{\parallel}) &\approx \lambda_{1}(s'-s'')|B|\cos{\left(2\phi_{S}+\phi_{B}+\phi_{0}\right)}\\
    &+\lambda_{2}\left((s')^2-(s'')^2\right)|B|\cos{\left(2\phi_{S}+\phi_{B}+\phi_{1}\right)}\\
    &+\lambda_{3}\left(s'-s''\right)^2|B|\cos{\left(-4\phi_{S}+\phi_{B}+\phi_{2}\right)},
\end{split}
\end{align}
for three general angles $\phi_{0},\phi_{1},\phi_{2}\in\left[0,\pi\right)$, and we find two new coupling coefficients $\lambda_{2}$ and $\lambda_{3}$. The angles $\phi_{0,1,2}$ are fixed by $R_{\pi}^{x,y}$ symmetries, as explained in the next subsection.

\subsection{Effects of $R_{\pi}^{x}$ and $R_{\pi}^{y}$ symmetries on the energy}
We now turn to explore further constraints imposed by the $R_{\pi}^{x}$ and $R_{\pi}^{y}$ symmetries. To do so, we must specify how the strain tensor changes due to layer flip, and reintroduce the valley filling dependence in the coupling coefficients $\lambda_{i}(\nu_{K},\nu_{K'})$. In order to account for both heterostrain and homostrain, we take $\chi=\pm 1$ to be the eigenvalue of the layer flip applied to the strain tensor, i.e. $\chi=-1$ ($\chi=+1$) for the heterostrain (homostrain).
The $R_{\pi}^{x}$ symmetry preserves the valleys and takes $\mathcal{S}\rightarrow\chi\mathcal{S}$, $\phi_{S}\rightarrow-\phi_{S}$ and $\phi_{B}\rightarrow-\phi_{B}$. For an energy in the form of Eq.~\eqref{eq:app_energy_only_c3}, we find that $\phi_{1}=\phi_{2}=0$, and that for heterostrain (homostrain) we have $\phi_{0}=\frac{\pi}{2}$ ($\phi_{0}=0$). We can combine both the case heterostrain and of homostrain into a single equation by writing:
\begin{align}\label{eq:app_energy_c3_r_pi_x}
\begin{split}
    E(\mathcal{S},\vec{B}_{\parallel}) &\approx
    \\
    &\left[\lambda_{1}(\nu_{K},\nu_{K'})(s'-s'')\cos{\left(2\phi_{S}+\phi_{B}+\frac{(1-\chi)\pi}{4}\right)}\right.\\
    &+\lambda_{2}(\nu_{K},\nu_{K'})\left((s')^2-(s'')^2\right)\cos{\left(2\phi_{S}+\phi_{B}\right)}\\
    &+\left.\lambda_{3}(\nu_{K},\nu_{K'})\left(s'-s''\right)^2\cos{\left(-4\phi_{S}+\phi_{B}\right)}\right]|B|,
\end{split}
\end{align}
and selecting the desired value of $\chi$.

Next, we consider the $R_{\pi}^{y}$ symmetry. This symmetry flips the valleys and takes $\mathcal{S}\rightarrow\chi\mathcal{S}$, $\phi_{S}\rightarrow \pi-\phi_{S}$ and $\phi_{B}\rightarrow\pi-\phi_{B}$. Changing the valleys is associated with a sign flip of the magnetization as seen by Eq.~\eqref{eq:energy_symmetry_c2}. This valley flip simply adds the rule $\lambda_{i}\rightarrow-\lambda_{i}$ for $i=1,2,3$. Using these transformation rules, one finds that  the $R_{\pi}^{y}$ symmetry does not impose any additional constraints, for both the heterostrained and homostrained TBG. Therefore, our final result is given by Eq.~\eqref{eq:app_energy_c3_r_pi_x}. The effects of $C_2$ and $\mathcal{T}$ are described in the main text. The magnetization immediately follows from this energy, as it is given to linear order in magnetic field:
\begin{align}\label{eq:app_magnetization_c3_r_pi_x}
\begin{split}
    M_{\parallel,x}+iM_{\parallel,y} &\approx
    -\lambda_{1}(\nu_{K},\nu_{K'})(s'-s'')e^{i(\chi-1)\pi/4}e^{-i2\phi_{S}}\\
    &-\lambda_{2}(\nu_{K},\nu_{K'})\left((s')^2-(s'')^2\right)e^{-i2\phi_{S}}\\
    &-\lambda_{3}(\nu_{K},\nu_{K'})\left(s'-s''\right)^2e^{i4\phi_{S}}.
\end{split}
\end{align}

\begin{figure}
    \centering
    \includegraphics[width=0.5\textwidth]{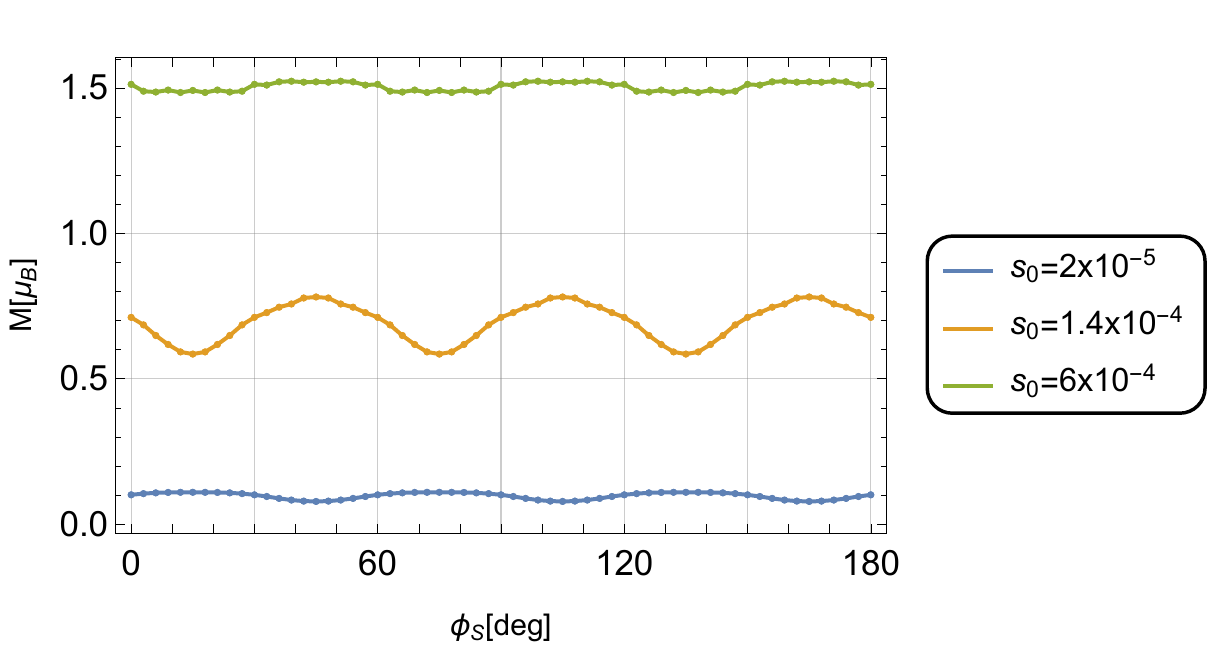}
    \caption{The valley-$K$ resolved magnitude of the magnetization \textit{vs} strain angle for the non-interacting TBG with $\theta=1.1^{\circ}$. For very weak strain the magnitude does not depend on strain. Increasing strain introduces angular dependence, and further increasing introduces higher order effects.}
    \label{fig:mag_abs_vs_phi}
\end{figure}

\begin{figure}
    \centering
    \includegraphics[width=0.5\textwidth]{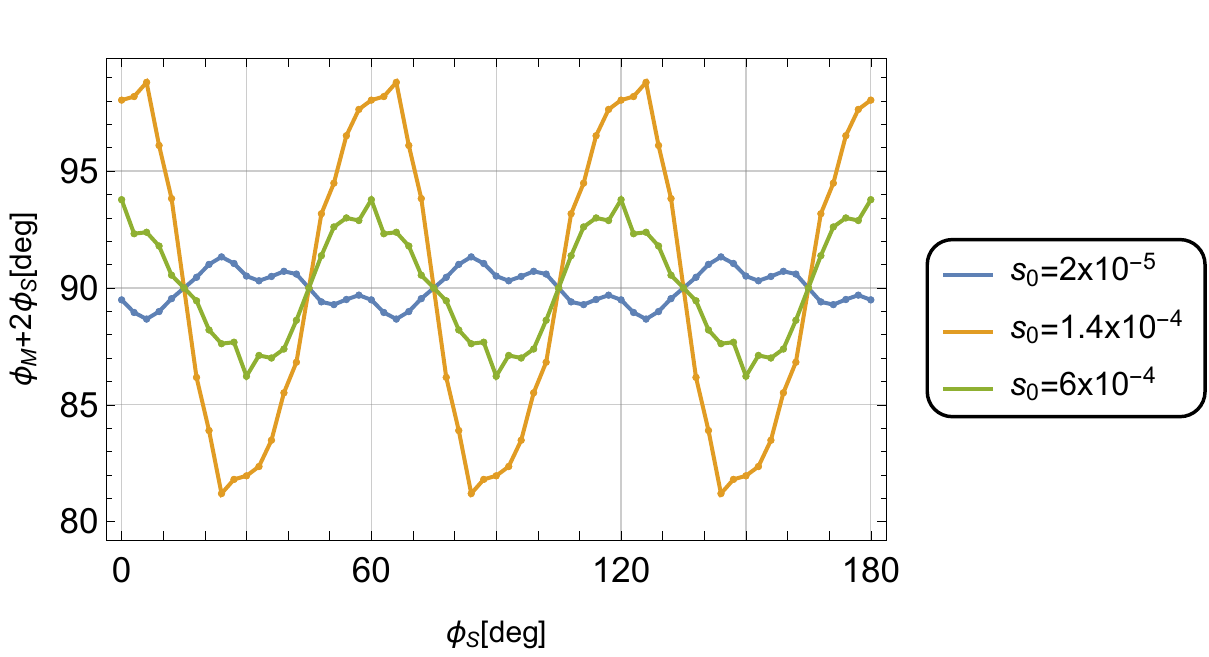}
    \caption{The valley-$K$ resolved magnetization orientation $\phi_{M}$ shifted by $2\phi_{S}$ as a function of strain angle $\phi_{S}$, for the non-interacting TBG with $\theta=1.1^{\circ}$. The oscillatory behaviour is a result of higher orders in strain.}
    \label{fig:mag_phi_vs_phi}
\end{figure}

The result in Eq.~\eqref{eq:app_magnetization_c3_r_pi_x} is important, as it tells us that the magnetization magnitude $\left|\vec{M}_{\parallel}\right|$ is no longer $\phi_{S}$ independent at second order in strain. 
There is no way for symmetry arguments to predict the amount of variance expected, nor when higher orders become significant. We can, however, plot the magnetization magnitude and orientation as a function of angle $\phi_{S}$ for a strain tensor with $s'=-s''=s_{0}$, and compare different values of $s_{0}$. For this choice of strain, the term with $\lambda_{2}$ drops, and we have two competing couplings - $\lambda_{0}$ and $\lambda_{1}$. The results are presented in Fig.~\ref{fig:mag_abs_vs_phi}, which displays how the magnetization magnitude varies with strain angle. For weak strain, the variation is small. Increasing the strain displays strong oscillations with frequency $6\phi_{S}$, in agreement with Eq.~ \eqref{eq:app_magnetization_c3_r_pi_x}. For even larger strain, higher order oscillations are observed, and the magnitude saturates. The orientation of the magnetization $\phi_{M}$ is plotted against the strain angle in Fig.~\ref{fig:mag_phi_vs_phi}. A linear phase of $2\phi_{S}$ is added to compensate for the rotation of the orientation due to the first order contribution. The oscillatory behaviour of the second order is clear even for weak strains, and the amplitude increases with strain until higher order effects become significant.


\section{Hartree-Fock Approximation}\label{sec:app_hartree_fock}

\subsection{The Hartree-Fock Hamiltonian}
We start by considering the complete interaction model. The interacting Hamiltonian $\hat{H}$ is given by a sum of a single-particle Hamiltonian $\hat{H}_{0}$ and an interaction term $\hat{V}$:
\begin{align}\label{eq:app_interacting_hamiltonian}
    \hat{H} &= \hat{H}_{0}+\hat{V},
    \\\label{eq:app_sp_hamiltonian}
    \hat{H}_{0} &= \sum_{\vec{k}}c_{\vec{k}}^{\dagger}\left(H_{\text{BM}}(\vec{k})-\Sigma^{HF}[\rho_{0}](\vec{k})\right)c_{\vec{k}},
    \\
    \hat{V} &= \frac{1}{2A}\sum_{\vec{q}}V(\vec{q}):\hat{n}_{\vec{q}}\hat{n}_{-\vec{q}}:,
\end{align}
where $c_{\vec{k}}^{\dagger}$ is a row vector of creation operators $c_{\vec{k},\alpha}^{\dagger}$ for Bloch eigenstates $\ket{u_{\vec{k},\alpha}}$ of $H_{BM}(\vec{k})$, which are labeled by the combined index $\alpha=(\tau,s,m)$ as defined in the main text. The total area of the system is denoted by $A$.

The single-particle Hamiltonian contains two parts: the BM Hamiltonian, $H_{BM}(\vec{k})$, and the Hartree-Fock self-energy of a reference state $\rho_0$,  $\Sigma^{HF}[\rho_0](\vec{k})$, which is subtracted to avoid double-counting, as discussed below. The BM Hamiltonian is given for valley $\tau=K$, 
by Eq.~\eqref{eq:h_bm_w_mag}, and for valley $\tau=K'$ by its time-reversal. The parameters in the BM Hamiltonian implicitly include interaction effects, which necessitates a subtraction scheme in order to avoid double counting. To this end, we deduct a Hartree-Fock self energy operator $\Sigma^{HF}[\rho_{0}](\vec{k})$ evaluated at a reference state with a single-particle density matrix $\rho_0$ that corresponds to two decoupled graphene layers \cite{Xie2020,Xie2021}. 
The resulting $\hat{H}_{0}$ is designed such that the Hartree-Fock Hamiltonian that corresponds to $\hat{H}_0+\hat{V}$ in the absence of inter-layer tunneling reproduces $H_{BM}$ with $t_{AA}=t_{AB}=0$. The density matrix $\rho_{0}$ is also responsible for deducting the background charge density, as it corresponds to a uniform system at charge neutrality.

The interaction term $\hat{V}$ is chosen to be a double-gated Coulomb potential, $V(\vec{q})=e^2/(2\epsilon\epsilon_0 q)\tanh(qd_{g})$ with $\epsilon_0$ being the vacuum permittivity, $d_{g}=40\rm{nm}$ being the distance to the gates. The dielectric constant is chosen to be $\epsilon=30$, for reasons to be explained in section~\ref{sec:app_hf_spectrum}. The operator $\hat{n}_{\vec{q}}$ is the electron density operator with wavevector $\vec{q}$, which can be written in the basis of the BM eigenstates as
\begin{equation}
    \hat{n}_{\vec{q}}
    =\sum_{\vec{k}}c_{\vec{k}}^{\dagger}\Lambda_{\vec{q}}(\vec{k})c_{\vec{k}+\vec{q}},
    \quad 
    \left[\Lambda_{\vec{q}}(\vec{k})\right]_{\alpha,\beta}=\braket{u_{\vec{k},\alpha}|u_{\vec{k}+\vec{q},\beta}}.
\end{equation}

The Hartree-Fock self energy operator $\hat{\Sigma}^{HF}$ for a given Slater determinant with a single-particle density matrix $\rho$ is given by:
\begin{align}\label{eq:app_hf_self_energy}
\begin{split}
    \hat{\Sigma}^{HF}[\rho](\vec{k}) &= \frac{1}{A}\sum_{\vec{g}}V(\vec{g}) c_{\vec{k}}^{\dagger}\Lambda_{\vec{g}}(\vec{k})c_{\vec{k}}\sum_{\vec{k}'}\tr{\rho(\vec{k}')\Lambda_{\vec{g}}^{*}(\vec{k}')}
    \\
    &-\frac{1}{A}\sum_{\vec{q}=\vec{g}+\vec{k}'}V(\vec{q})
    c_{\vec{k}}^{\dagger}\Lambda_{\vec{q}}(\vec{k})\rho^{T}(\vec{k}+\vec{q})\Lambda_{\vec{q}}^{\dagger}(\vec{k})c_{\vec{k}},
\end{split}
\end{align}
where we adopted a periodic gauge, i.e. $c_{\vec{k}+\vec{g}}^{\dagger} = c_{\vec{k}}^{\dagger}$ for any moir\'{e} reciprocal lattice vector $\vec{g}$. The wavevector $\vec{k}'$ is limited to the first BZ, while $\vec{q}$ is any wavevector compatible with the boundary conditions. The first and second terms in Eq.~\eqref{eq:app_hf_self_energy} correspond to the Hartree and Fock self energies, respectively. In this work we neglect any flavor coherence, and Bloch wavefunctions are assumed orthogonal between different flavors. This sets all form factors, interpolated density matrices and their corresponding self energies to be diagonal in flavor.

The Hartree-Fock procedure is an approximation method for obtaining the ground state of \eqref{eq:app_interacting_hamiltonian}. The Hartree-Fock Hamiltonian presented in Eq. \eqref{eq:schf_def} can be rewritten in terms of $\hat{H}_{BM}$ and $\hat{\Sigma}^{HF}$ as:
\begin{equation}\label{eq:app_hf_se_subtracted}
    \hat{H}^{HF} = \hat{H}_{BM}+\hat{\Sigma}^{HF}[\rho-\rho_{0}].
\end{equation}
The purpose of our interpolation scheme is to provide an approximated form for $\hat{\Sigma}^{HF}$ without self-consistently determining $\rho$. Each interpolation configuration in Table~\ref{tab:flavor_interpolation} fixes the density matrix $\rho(\nu)$ in terms of the filling, as defined in Eq. \eqref{eq:density_interpolation_with_flavors}. 

\subsection{Quasi-particle excitation spectrum}\label{sec:app_hf_spectrum}

\begin{figure}
    \centering
    \includegraphics[width=0.5\textwidth]{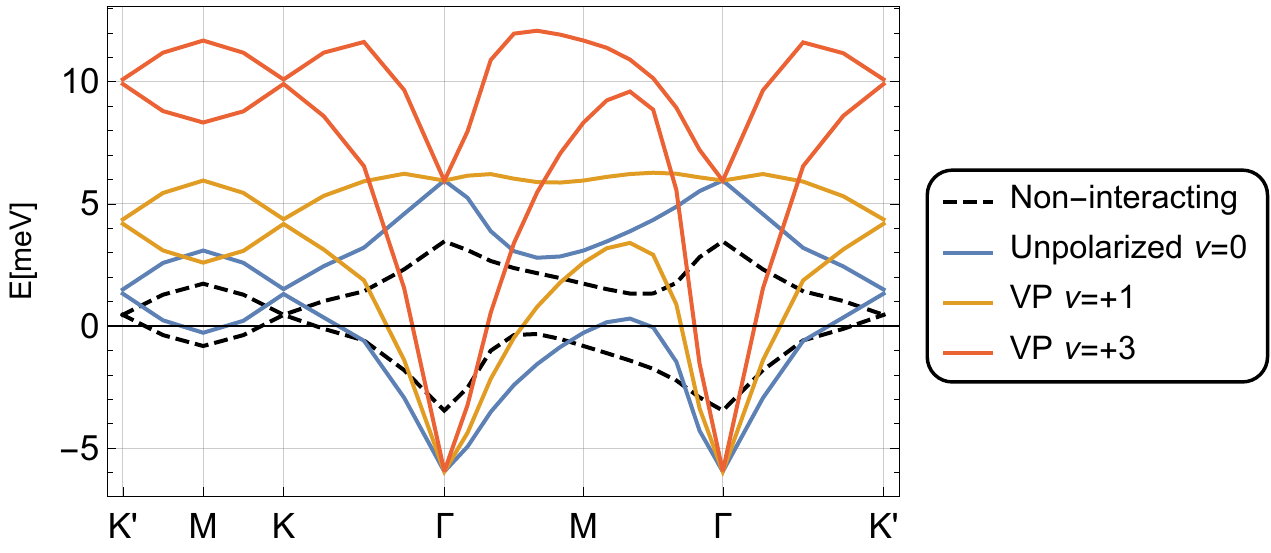}
    \caption{Hartree-Fock flat band spectrum of flavor $(K,\uparrow)$ and $\theta=1.1^{\circ}$ for different interpolation configurations. The non-interacting spectrum is slightly particle-hole asymmetric, leading to a deviation of the Dirac points from zero energy. The rest of the plots are shifted in energy such that at the $\Gamma$ point the two bands sit at exactly opposite energies.}
    \label{fig:hf_flat_bands_var_interpolation}
\end{figure}

In Fig.~\ref{fig:hf_flat_bands_var_interpolation} we plot the flat bands of the Hartree-Fock spectrum of Eq.~\eqref{eq:schf_def} for different interpolation configurations. We compare between the non-interacting spectrum of the unpolarized configuration at $\nu=0$, and the valley polarized configurations $\nu=+1$ and $\nu=+3$. We note that the increase in Dirac velocity is a consequence of the Fock self energy, and the relative shift between the Dirac points and the $\Gamma$ point is due to the Hartree self energy, as described in Ref.~[\onlinecite{Xie2021}]. 

\begin{figure}
    \centering
    \includegraphics[width=0.5\textwidth]{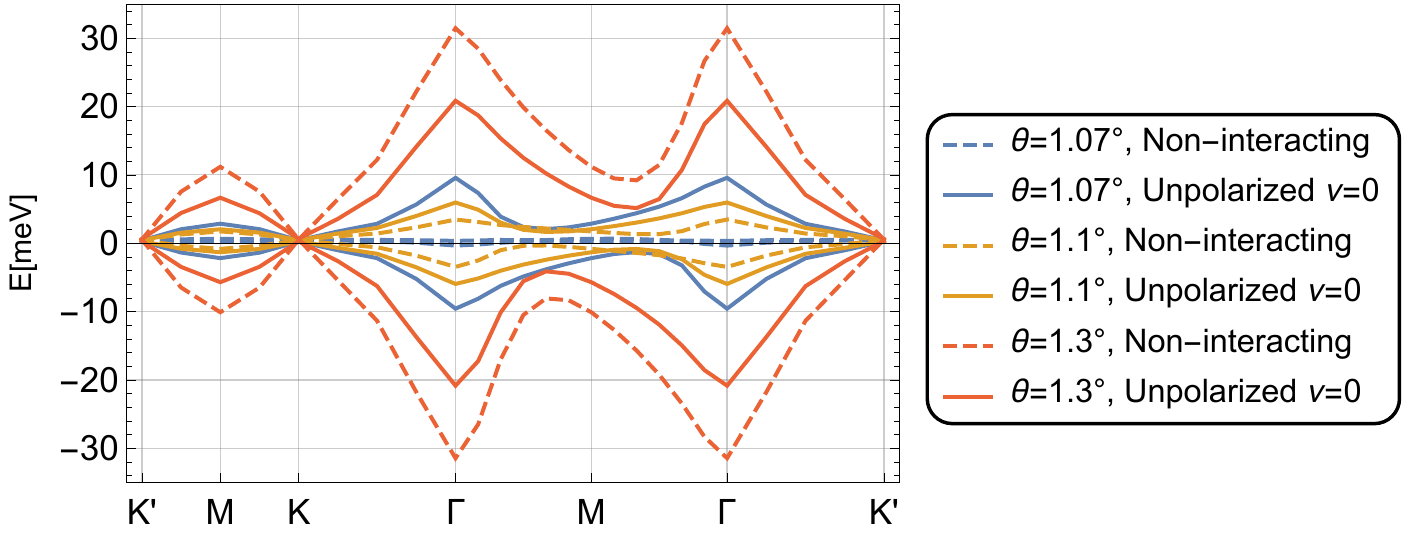}
    \caption{Hartree-Fock flat band spectrum for different twist angles. There is a correspondence between bandwidth (or Dirac velocity) and the sensitivity to strain, implying the effect of interactions on magnetization is mostly determined by the Fock term bandwidth renormalization.}
    \label{fig:hf_flat_bands_var_twist}
\end{figure}

The dielectric constant $\epsilon$ is treated as a phenomenological parameter that includes both the screening due to the substrate and the remote bands. We fix it by requiring that the obtained magic angle roughly matches the experimental one. We expect a minimal bandwidth for $\theta\sim1.07^{\circ}-1.1^{\circ}$, and a wider one for $\theta\sim1.3^{\circ}$. This is accomplished by our choice of $\epsilon=30$, for which the bandwidths are presented in the quasi-particle excitation spectrums in Fig.~\ref{fig:hf_flat_bands_var_twist}. Note that for this choice the band is actually narrower for $\theta=1.1^{\circ}$ than for $\theta=1.07^{\circ}$, and that the interaction decreases the bandwidth for $\theta=1.3^{\circ}$. This figure therefore complements Fig.~\ref{fig:mag_vs_strain_int} in the main text by demonstrating how the bandwidth (or Dirac velocity) is related to the magnetization response.

\section{Reduction to a single flavor}\label{sec:app_single_flavor}
As described in the main text, the in-plane magnetization in Eq.~\eqref{eq:mag_from_chem_pot} for the valley-polarized states can be computed by focusing on a single flavor block in the Hartree-Fock interpolation. However, as seen in Eq.~\eqref{eq:app_hf_self_energy}, the Hartree term consists of a sum over charge contributions from all the flavors. Therefore, in the general case, all the flavor blocks in the density matrices $\rho,\rho_{0}$ are required to obtain the self energy for a specific flavor. However, using our interpolation scheme, we can simplify the computation and calculate the Hartree term with a single flavor block. 

Consider the following term that appears in the Hartree self energy:
\begin{equation}
    \braket{n_{-\vec{g}}}_{\rho-\rho_{0}}=\sum_{\vec{k}'}\tr{\left(\rho(\nu)-\rho_{0}(\vec{k}')\right)\Lambda_{\vec{g}}^{*}(\vec{k}')},
\end{equation}
where the trace is taken over the spin, valley, and band indices. Since both the density matrices and the form factors are diagonal in flavor, we can split the trace into a sum over flavor sectors denoted by a subscript $(\tau,s)$ for valley and spin:
\begin{align}
\begin{split}
    \braket{n_{-\vec{g}}}_{\rho-\rho_{0}}
    &=\sum_{\vec{k}'}
    \sum_{\tau,s}\tr{\left(\rho_{\tau,s}(\nu_{\tau,s})-\rho_{0;\tau,s}(\vec{k}')\right)\Lambda_{\vec{g}}^{*}(\vec{k}')}
    \\
    &= \sum_{\tau,s}
    \left(\braket{n_{-\vec{g}}}_{\rho;(\tau,s)}-\braket{n_{-\vec{g}}}_{\rho_{0};(\tau,s)}\right),
\end{split}
\end{align}
where here the trace is over the band indices. 

For the density matrix $\rho_{0;\tau,s}(\vec{k})$ the expression above is simple: there is no charge at any nonzero wavevector, and all flavors are symmetric, $\braket{n_{-\vec{g}}}_{\rho_{0};(\tau,s)}=n_{0}\delta_{\vec{g}=\vec{0}}$. For the interpolated density matrix $\rho_{\tau,s}(\nu_{\tau,s})$, each flavor is a combination of density matrices of filled and empty flavors. Suppose that all flavors are filled to $\nu_{\tau,s}=\nu_{0}$, such that the flavor blocks in the density matrix are related by symmetry. The spin $SU(2)$ symmetry implies that $\braket{n_{-\vec{g}}}_{\rho;(\tau,\uparrow)}=\braket{n_{-\vec{g}}}_{\rho;(\tau,\downarrow)}$. The $C_{2}$ symmetry (which holds in the presence of strain) implies that $\braket{n_{-\vec{g}}}_{\rho;(K,s)}=\braket{n_{+\vec{g}}}_{\rho;(K',s)}$. Together with $\hat{n}_{-\vec{g}}=\hat{n}_{+\vec{g}}^{\dagger}$ we find that $\braket{n_{-\vec{g}}}_{\rho;(K,s)}=\braket{n_{-\vec{g}}}_{\rho;(K',s)}^{*}$. 
The last symmetry we invoke is the anti-unitary $C_{2}\mathcal{T}=\sigma_{x}\mathcal{K}$, where $\sigma_{x}$ flips the sublattice and $\mathcal{K}$ is the complex conjugation operator. This symmetry implies that $\ket{u_{\vec{k},\alpha}}=e^{i\phi_{\vec{k},\alpha}}\sigma_{x}(\ket{u_{\vec{k},\alpha}})^{*}$, with a gauge dependent phase $\phi_{\vec{k},\alpha}$. In the periodic gauge it is easy to show that we have $\phi_{\vec{k}+\vec{g},\alpha}=\phi_{\vec{k},\alpha}$ for any lattice vector $\vec{g}$. This symmetry immediately implies that the diagonal elements of the form factors, $\left[\Lambda_{\vec{g}}(\vec{k}')\right]_{\alpha,\alpha}$, are strictly real. Since $\rho_{\tau,s}(\nu_{\tau,s})$ is diagonal as well, we find that $\braket{n_{-\vec{g}}}_{\rho;(\tau,s)}$ is strictly real, hence $\braket{n_{-\vec{g}}}_{\rho;(K,s)}=\braket{n_{-\vec{g}}}_{\rho;(K',s)}$.

Combining the arguments above, we find that $\rho_{K,\uparrow}$ and $\rho_{0;K,\uparrow}$ can be used to calculate any flavor's contribution to the overall charge density by using them with each flavor's respective filling:
\begin{align}\label{eq:app_charge_density_combined_flavor}
\begin{split}
    \braket{n_{-\vec{g}}}_{\rho-\rho_{0}} 
    &=
    \sum_{\vec{k}'}\sum_{\tau,s}\tr{\left(\rho_{K,\uparrow}(\nu_{\tau,s})-\rho_{0;K,\uparrow}(\vec{k}')\right)\Lambda_{\vec{g}}^{*}(\vec{k}')}.
\end{split}
\end{align}
Eq.~\eqref{eq:app_charge_density_combined_flavor} implies that the Hartree-Fock self energy for the flavor $(\tau,s)=(K,\uparrow)$ in Eq.~\eqref{eq:app_hf_se_subtracted} can be written solely with $\rho_{K,\uparrow}$ and $\rho_{0;K,\uparrow}$ by using the following effective density matrix for the Hartree and Fock terms separately:
\begin{align}\label{eq:app_hf_using_single_flavor}
\begin{split}
    \Sigma^{HF}[\rho-\rho_{0}]_{K,\uparrow}
    &= \Sigma^{H}\left[\left(\Sigma_{\tau,s}\rho_{K,\uparrow}(\nu_{\tau,s})\right)-4\rho_{0;K,\uparrow}(\vec{k}')\right]
    \\
    &+\Sigma^{F}[\rho_{K,\uparrow}(\nu_{K,\uparrow})-\rho_{0;K,\uparrow}(\vec{k}')].
\end{split}
\end{align}
We therefore restrict all density matrices in Appendix~\ref{sec:app_numerical} to be in the $K\uparrow$ flavor block, and solve the Hartree-Fock Hamiltonian for this flavor.

\section{Numerical procedure for calculating the chemical potential}\label{sec:app_numerical}
The purpose of this appendix is to detail technical aspects involving calculating the chemical potential as a function of filling and its gradient with respect to magnetic field. 

We start by detailing the numerical parameters used in the computation. The non-interacting BM Hamiltonian of flavor $(\tau,s)=(K,\uparrow)$ is constructed symmetrically around the $\Gamma$ point in the moir\'{e} BZ, with the momentum space cutoff taken to include all hopping sites within a hexagon of edge size of $3\sqrt{3}[K\theta]$, i.e. an edge size of $R_{BZ}=3$ in units of a reciprocal lattice wavevector magnitude. This cutoff includes 27 hopping sites in each layer. Multiplying by the sublattice degree of freedom, we obtain $N_{\text{bands}}=108$ bands for each quasimomentum point in the moir\'{e} BZ. Note that, by construction, this Hamiltonian is not exactly $C_3$ symmetric about any point beside the $\Gamma$ point. The quasimomentum sampling grid within the first BZ is centered around the $\Gamma$ point with the number of unit cells taken to be $N_{u.c.}=133$. The grid is sampled first along the unstrained moir\'{e} lattice vectors, and then transformed under the desired strain, in order to preserve $N_{u.c.}$ across different strains.  

The chemical potential for a particular filling factor $\mu(\nu)$ is calculated from the spectrum of the HF Hamiltonian in Eq.~\eqref{eq:schf_def} using the procedure outlined in the main text. We will now describe this procedure with further technical details. The first step is to use the reduction scheme in Eq.~\eqref{eq:app_hf_using_single_flavor} to restrict our attention of the HF Hamiltonian to a single flavor block. Next, we simplify the diagonalization of this Hamiltonian by projecting the self energy $\hat{\Sigma}^{HF}[\rho](\vec{k})$ to the two flat bands in the non-interacting Hamiltonian, as in Ref.~[\onlinecite{Xie2021}]. We emphasize that all the remote bands are accounted for by considering the density matrix $\rho(\vec{k})$ and the form factors $\Lambda_{\vec{g}}(\vec{k})$ to be the entire $N_{\text{bands}}\times N_{\text{bands}}$ matrices.  The flat-band projection truncates the Hartree Fock operator in Eq.~\eqref{eq:schf_def} to be a $2\times 2$ matrix for each quasimomentum. 

After projection, the two Hartree-Fock bands consist of $2N_{u.c.}$ energy eigenvalues from all quasimomenta in the moir\'{e} BZ combined. The filling factor range $\nu\in(\nu_{e},\nu_{f})$ is discretized accordingly to $\nu(m)$ where $m= 1,2,\dots, 2N_{u.c.}$. Starting with the lowest filling $\nu(1)=\nu_{e}$, we compute the \textit{sorted} energy eigenvalues $\epsilon_{i}(\nu(1))$, $i= 1,2,\dots, 2N_{u.c.}$, such that $\epsilon_{i}<\epsilon_{i+1}$.  The chemical potential $\mu(\nu=\nu(1))$ is given by choosing the lowest energy eigenvalue $\epsilon_{1}$. We iterate this procedure over $m$ by computing the density matrix at the next filling factor $\nu(m)$, computing its corresponding projected HF Hamiltonian, sorting its eigenvalues and selecting the new $m$-th energy $\epsilon_{m}$. In summary, the chemical potential $\mu(\nu)$ is a list of $2N_{u.c.}$ values, whose $m$-th entry corresponds to the $m$-th eigenvalue of a HF Hamiltonian interpolated to the $m$-th filling. We note that the information about the specific value of $\vec{k}$ which is used for calculating $\mu(\nu)$ is lost. 

For the purpose of calculating the magnetization we computed the chemical potential at four magnetic fields,
$\vec{B}_{\parallel}\in\lbrace -B\hat{x},-B\hat{y},B\hat{x},B\hat{y}\rbrace$, and extracted the gradient of the chemical potential $\mu(\nu,\vec{B}_{\parallel})$ at $\vec{B}_{\parallel}=0$. Next, one might expect that integrating over filling would yield the desired magnetization, as Eq.~\eqref{eq:mag_from_chem_pot} requires.
Unfortunately, there is one critical problem with immediately performing this integration. The problem is most clearly explained for TBG without strain. In the absence of strain, we expect the magnetization to be exactly zero as a function of filling. Since quasi-particle states are filled one at a time, the $C_{3}$ symmetry is explicitly broken. We therefore find that the magnetization per moir\'{e} unit cell varies as a function of filling, reaching magnitudes of the order of $1\mu_{B}/N_{u.c.}$ that are cancelled only after filling all three $C_{3}$-symmetric states. The problem persists at weak strains, but becomes irrelevant once all single-particle magnetization contributions are aligned. 

To remedy this we introduce a finite temperature to smooth out the chemical potential. Given a filling $\nu$ and temperature $T$, the chemical potential is set to fix the Fermi-Dirac distribution $f_{FD}$ to the number of particles $N_{\text{p}}(\nu)$:
\begin{equation}
    N_{\text{p}}(\nu) = \sum_{\vec{k}}f_{FD}\left(\epsilon_{\vec{k}}(\nu,\vec{B}_{\parallel})-\mu(\nu,\vec{B}_{\parallel})\right),
\end{equation}
where $\epsilon_{\vec{k}}(\nu,\vec{B}_{\parallel})$ are the HF eigenvalues for the density matrix corresponding to filling $\nu$, evaluated in the presence of a magnetic field $\vec{B}_{\parallel}$.
Taking the gradient with respect to magnetic field, we find the self consistent equation for $\mu(\nu,\vec{B}_{\parallel})$:
\begin{equation}
    \frac{\partial \mu}{\partial \vec{B}_{\parallel}}
    =\left(\sum_{\vec{k}}\frac{\partial f_{FD}}{\partial \epsilon_{\vec{k}}}\right)^{-1}
    \sum_{\vec{k}}\frac{\partial f_{FD}}{\partial \epsilon_{\vec{k}}} 
    \frac{\partial \epsilon_{\vec{k}}}{\partial \vec{B}_{\parallel}}.
\end{equation}
We use the following equation as a definition for the ``smoothed" chemical potential gradient at $\vec{B}_{\parallel}=0$, and evaluate it using the chemical potential $\mu(\nu,\vec{B}_{\parallel}=0)$. This chemical potential gradient is the one used for all the numerical results. The specific temperature $T$  varies for each plot but is of the order of average level spacing, i.e. $T\sim E_{BW}/N_{u.c.} \sim 0.1\rm{meV}$, where $E_{BW}$ is the combined bandwidth of the HF flat bands.

We have verified that our results for magnetization magnitudes $|M_{\parallel}|\gtrsim 0.1 \mu_{B}$ are sufficiently converged for our choice of $R_{BZ}$, $N_{u.c.}$, and stable in a range of temperatures $T$. For lower magnetization values we have recovered the linear dependence of the magnetization vs. strain, but the precise slope has a small dependence on the numerical parameters.

\end{document}